\documentclass[aps,groupedaddress]{revtex4}
\usepackage{epsfig}
\usepackage{graphicx}
\usepackage[T1]{fontenc}
\usepackage{ae}
\usepackage{color}
\usepackage[latin1]{inputenc}
\usepackage{amssymb,amsbsy,amsmath}
\usepackage{bbm}

\begin{document}



\title[Rabi Geometry]
      {On the geometry of the classical Rabi problem}
\author{Heinz-J\"urgen Schmidt
}
\address{  Universit\"at Osnabr\"uck,
Fachbereich Physik,
 D - 49069 Osnabr\"uck, Germany}


\begin{abstract}
We investigate the motion of a classical spin precessing around a periodic magnetic field using
Floquet theory as well as elementary differential geometry and considering a couple of examples.
Under certain conditions the r\^{o}le of spin and magnetic
field can be interchanged, leading to the notion of ``duality of loops" on the Bloch sphere.
\end{abstract}

\maketitle

\section{Introduction}\label{sec:I}

The Rabi problem  usually refers to the response of an atom to an applied harmonic electric field,
with an applied frequency very close to the atom's natural frequency \cite{R37}, \cite{Shirley65}.
The corresponding classical model to describe this problem would
be, for example, a driven, damped harmonic oscillator. In this paper, however, we will understand by the ``classical Rabi problem"
a different approach: The atom can be approximated by a two-level system such that its semi-classical Hamiltonian assumes the form of 
Zeeman term in an  $s=1/2$ spin system:
\begin{equation}\label{I1}
 H=\omega_0\,S_z + h_1(t)\,S_x+ h_2(t)\,S_y+h_3(t)\,S_z
 \;,
\end{equation}
where the $S_x,S_y,S_z$ are the  $s=1/2$ spin operators. If $\psi(t)$ is a solution of the corresponding Schr\"odinger equation
\begin{equation}\label{I1a}
  {\sf i}\hbar \frac{d}{dt}\psi(t)=H\,\psi(t)
  \;,
\end{equation}
then the projector $P(t)=\left|\psi(t)\rangle\langle \psi(t)\right|$ can be expanded as a linear combination of the spin operators:
\begin{equation}\label{I2}
P(t)= \frac{1}{2}\,{\mathbbm 1}+ {\mathbf s}_1(t)\,S_x+{\mathbf s}_2(t)\,S_y+{\mathbf s}_3(t)\,S_z
\;.
\end{equation}
It follows that ${\mathbf s}(t)\equiv \left({\mathbf s}_1(t),{\mathbf s}_2(t),{\mathbf s}_3(t)\right)^\top$
will be a unit vector that obeys the same equation of motion
\begin{equation}\label{I3}
\frac{d}{dt}{\mathbf s}(t)={\mathbf h}(t)\times {\mathbf s}(t)
\end{equation}
as a classical magnetic moment performing a Larmor precession around the time dependent periodic magnetic field
${\mathbf h}(t)\equiv \left(h_1(t),h_2(t),h_3(t)+\omega_0\right)^\top$.
The study of this equation will be called the ``classical Rabi problem" in what follows.

According to the preceding remarks it seems that the solutions of (\ref{I3}) provide the information
about the corresponding solutions of (\ref{I1a}) up to a (time-depending) phase factor. But it can be shown \cite{S18} that
also this phase factor can be reconstructed from periodic solutions of (\ref{I3}) by means of certain integrals.
Here we encounter the rare case where a quantum problem and the corresponding classical problem are essentially equivalent.
This endows the classical Rabi problem with additional importance concerning quantum applications.

The differential equation (\ref{I3}) can be explicitly solved only in a few cases of physical interest,
the most prominent one being a constant field superimposed by a monochromatical, circularly
polarized field perpendicular to the constant one \cite{R37}. The analogous problem with a linearly polarized
field component is solvable in terms of confluent Heun functions, see \cite{MaLi07}, \cite{XieHai10} and
\cite{SSH19} for the corresponding $s=1/2$ Schr\"odinger equation. In this paper we will shift the problem
of finding solutions of (\ref{I3}) to the study of geometric relations between such solutions and to the interplay
between Floquet theory, differential geometry of the unit sphere and duality of loops. Not all results will be new,
but we will provide new proofs that only use properties of solutions of the classical Rabi problem that are easier to visualize
and do not resort to the mathematics of the underlying Schr\"odinger equation.

The structure of the paper will be as follows. After general remarks on the classical Rabi problem in section \ref{sec:G}
we will in the main part (section \ref{sec:M}) specialize to periodic driving. The existence of periodic solutions of (\ref{I3})
is shown in subsection \ref{sec:P}. This result can be re-phrased in term of Floquet theory, see subsection \ref{sec:F}, where the
central notion of the ``quasienergy" is introduced. The ``geometric part" of the quasienergy can be written as an integral
involving the geodesic curvature of the periodic solution of (\ref{I3}) and, by virtue of the theorem of Gauss-Bonnet, will be related
to the solid angle enclosed by this solution, see subsection \ref{sec:QI}.
Alternatively, the geometric part of the quasienergy can be related to the phase shift of a second solution of (\ref{I3}) that
is orthogonal to the periodic solution if the magnetic field is replaced by an equivalent ``dual field", see subsection \ref{sec:D}.
This confirms the well-known connection between the Rabi problem and the geometric phase introduced by M.~V.~Berry and others.
The mentioned duality between spin loops and magnetic loops is further elaborated in subsection \ref{sec:D}. It can be used
to generate new classes of solutions, see subsection \ref{sec:N}. Finally, we check the agreement between the integral representations
for the quasienergy obtained in this paper and previous formulations, see subsection \ref{sec:QII}, and close with a summary.

\section{Generalities}\label{sec:G}
We consider the equation of motion
\begin{equation}\label{G1}
  \frac{d}{d\tau}{\mathbf s}(\tau)={\mathbf h}(\tau)\times {\mathbf s}(\tau)
\end{equation}
describing the time evolution of a classical spin vector ${\mathbf s}(\tau)$ due to the precession
around the time-dependent magnetic field
\begin{equation}\label{G1a}
 {\mathbf h}(\tau)=\left(
 \begin{array}{c}
   h_1(\tau) \\
   h_2(\tau) \\
   h_3(\tau)
 \end{array}
 \right)
 \;,
\end{equation}
measured in units of (Larmor) frequency. In the following we will use $\tau$ for the time variable and $\sigma$ for another arbitrary
parameter that is a function of $\tau$. Differentiation w.~r.~t.~$\tau$ will be denoted by a dot.
The arc length parameter of certain curves described by the spin vector or by the magnetic field vector
will be denoted by $s$ or $r$, resp..

Since the scalar product ${\mathbf s}\cdot{\mathbf s}=\|{\mathbf s}\|^2$ is conserved under (\ref{G1}) one usually assumes that
\begin{equation}\label{G2}
  \|{\mathbf s}(\tau)\|=1
  \;,
\end{equation}
that is, ${\mathbf s}(\tau)$ is moving on the (unit) Bloch sphere.

\begin{figure}[t]
\centering
\includegraphics[width=0.7\linewidth]{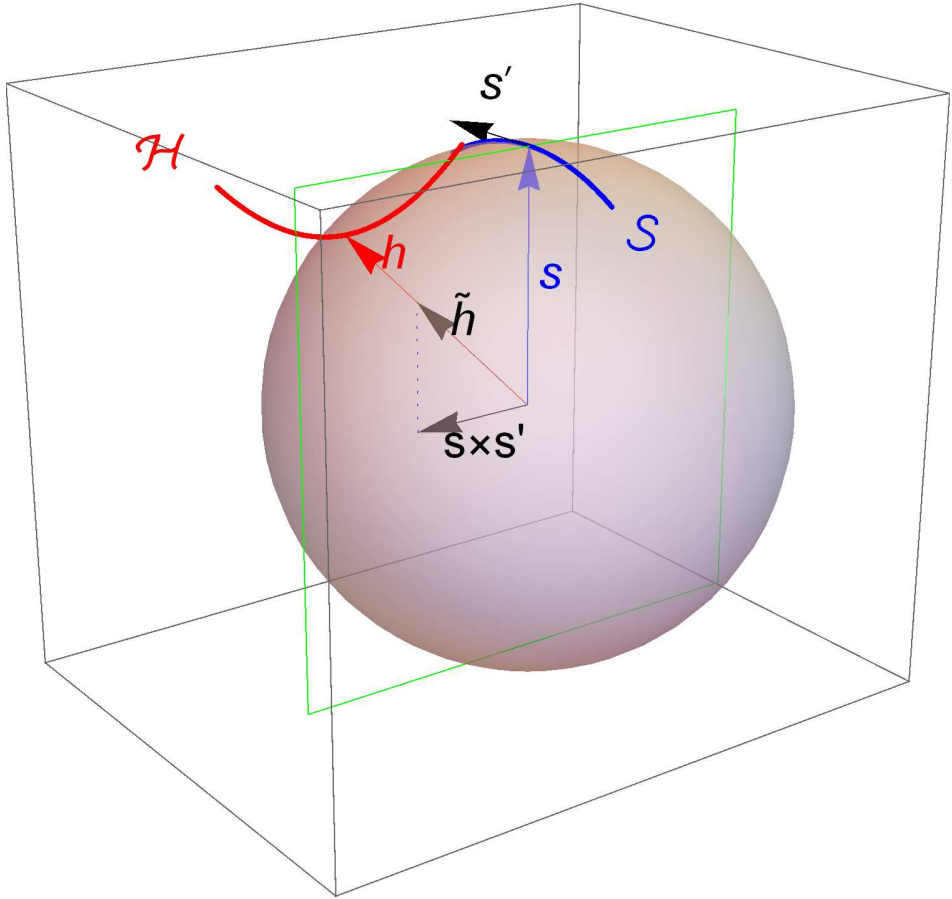}
\caption{Illustration of the construction described in the text. The (blue) spin vector ${\mathbf s}\in{\mathcal S}$
and its tangent vector ${\mathbf s}'$ define the plane ${\mathbf s}'^\perp$ indicated by green lines.
The (red) field vector ${\mathbf h}$ is found as the intersection of  ${\mathbf s}'^\perp$ and ${\mathcal H}$.
It must be proportional to $\tilde{\mathbf h}={\mathbf s}\times {\mathbf s}'+\lambda {\mathbf s}$
which yields the original $\tau$-parametrization by means of (\ref{G11}) and  (\ref{G12}).
}
\label{FIGSBR}
\end{figure}

Interestingly, the inverse problem of finding ${\mathbf h}(\tau)$ if the parametrized curve
${\mathbf s}(\tau)$ on the Bloch sphere is given, has the elementary solution
\begin{equation}\label{G3}
{\mathbf h}(\tau)={\mathbf s}(\tau)\times \dot{\mathbf s}(\tau) +\lambda(\tau) {\mathbf s}(\tau)
\;,
\end{equation}
where $\lambda(\tau)$ is an arbitrary smooth function.
In order to prove (\ref{G3}) we consider (\ref{G1}) as an inhomogeneous, linear equation for the unknown ${\mathbf h}(\tau)$
if ${\mathbf s}(\tau)$ and $\dot{\mathbf s}(\tau)$ are given for any time $\tau$ such that ${\mathbf s}(\tau)\cdot\dot{\mathbf s}(\tau)=0$.
A special solution of (\ref{G1}) is given by
\begin{equation}\label{G4}
 {\mathbf h}^{(0)}(\tau)={\mathbf s}(\tau)\times \dot{\mathbf s}(\tau)
 \;,
\end{equation}
since
\begin{eqnarray}
\label{G5a}
  {\mathbf h}^{(0)}\times {\mathbf s} &=& \left( {\mathbf s}\times \dot{\mathbf s}\right)\times{\mathbf s}   \\
  \label{G5b}
   &=& \underbrace{{\mathbf s}\cdot{\mathbf s}}_1\;\dot{\mathbf s} -\underbrace{ {\mathbf s}\cdot\dot{\mathbf s}}_0\;{\mathbf s}\\
    \label{G5c}
   &=& \dot{\mathbf s}
   \;.
\end{eqnarray}
The homogenous equation corresponding to (\ref{G1}) reads
\begin{equation}\label{G6}
  {\mathbf 0}={\mathbf h}(\tau)\times {\mathbf s}(\tau)
\end{equation}
and has the general solution
\begin{equation}\label{G7}
  {\mathbf h}(\tau)=\lambda(\tau){\mathbf s}(\tau)
  \;.
\end{equation}
Adding ${\mathbf h}^{(0)}(\tau)$ and (\ref{G7}) yields the general solution (\ref{G3}) of the inverse problem (\ref{G1}).

Next we will consider rather arbitrary parametrizations of the curves described by ${\mathbf s}$ and ${\mathbf h}$
that are given by (locally) smooth $1:1$ functions $\tau\mapsto \sigma(\tau)$ the (local) inverse denoted in a
somewhat sloppy but usual way by $\sigma\mapsto \tau(\sigma)$.
Conceptually, this means that we pass from the \textit{parametric curve} given by $\tau\mapsto {\mathbf s}(\tau)$ to the \textit{curve} ${\mathcal S}$
on the unit Bloch sphere without singling out a particular parametrization.
${\mathcal S}$ can be defined as the image of the map ${\mathbf s}:{\mathbbm R}\rightarrow S^2,\;\tau\mapsto {\mathbf s}(\tau)$.
Analogously, ${\mathcal H}$ will be the curve swept by the magnetic field without assuming any special parametrization.

The equation of motion (\ref{G1}) is transformed under the local parameter change $\tau\mapsto \sigma(\tau)$ as follows:
\begin{equation}\label{G8}
 \frac{d}{d\sigma}{\mathbf s}(\tau(\sigma))= \frac{d\tau}{d\sigma}\frac{d}{d\tau}{\mathbf s}(\tau(\sigma))
 =\frac{d\tau}{d\sigma}{\mathbf h}(\tau(\sigma))\times {\mathbf s}(\tau(\sigma))
 \;.
\end{equation}
This means that the transformed equation of motion has the form
\begin{equation}\label{G9}
   \frac{d}{d\sigma}{\mathbf s}(\sigma)=\tilde{\mathbf h}(\sigma)\times {\mathbf s}(\sigma)
\end{equation}
with a modified magnetic field $\tilde{\mathbf h}(\sigma)=\frac{d\tau}{d\sigma}{\mathbf h}(\tau(\sigma))$ that has the
same direction as the original field but possibly a different length. In case of $\frac{d\tau}{d\sigma}<0$,
$\tilde{\mathbf h}(\sigma)$ may even point into the opposite direction of ${\mathbf h}(\tau(\sigma))$ but
it is still ``projectively equivalent" to ${\mathbf h}(\tau(\sigma))$.

As an application of the preceding equations we consider the constant magnetic field
\begin{equation}\label{G9a}
  {\mathbf h}=\left(
  \begin{array}{c}
    0 \\
    0 \\
    F
  \end{array}
  \right)
  \;,
\end{equation}
and the corresponding elementary solution of (\ref{G1}) describing a precessing spin with constant angular velocity $F$
and forming an angle $\theta=\arccos z,\;-1<z<1$ with the magnetic field:
\begin{equation}\label{G9b}
  {\mathbf s}(\tau)=\left(
  \begin{array}{c}
    \sqrt{1-z^2} \cos F \tau\\
     \sqrt{1-z^2} \sin F \tau \\
   z
  \end{array}
  \right)
  \;.
\end{equation}
The parameter change
\begin{equation}\label{G9c}
 \tau=\frac{1}{\omega}\sin \omega \sigma\quad \Rightarrow\quad \frac{d\tau}{d\sigma}=\cos\omega \sigma
 \;,
 \end{equation}
gives rise to a modified field
\begin{equation}\label{G9d}
  \tilde{\mathbf h}(\sigma)=\frac{d\tau}{d\sigma}\,{\mathbf h}=
  \left(
  \begin{array}{c}
    0 \\
    0 \\
   F\,\cos\omega \sigma
  \end{array}
  \right)
  \;,
\end{equation}
such that the spin vector ${\mathbf s}$ can be written as a function of the new parameter $\sigma$, setting $f\equiv \frac{F}{\omega}$,
\begin{equation}\label{G9e}
 {\mathbf s}(\sigma)=\left(
  \begin{array}{c}
    \sqrt{1-z^2}\, \cos\left( f \sin\omega \sigma\right)\\
     \sqrt{1-z^2}\, \sin \left( f \sin\omega \sigma\right) \\
   z
  \end{array}
  \right)
\end{equation}
and satisfies the transformed equation of motion (\ref{G9}),
as can be easily confirmed by direct computation.
This solution has been used to obtain results for the case of a monochromatic, linearly polarized magnetic field in the limit
of a vanishing constant field component, see, e.~g., \cite{S18}.

We note the curiosity that the system $({\mathcal S},{\mathcal H})$ of two curves that is derived from
a solution of (\ref{G1}) in the way described above can be considered as a kind of ``clock"
in the sense that it allows, at least locally, the reconstruction of the original parametrization
$\tau\mapsto {\mathbf s}(\tau)$ and $\tau\mapsto {\mathbf h}(\tau)$. To show this we start with an arbitrary local parametrization
$ {\mathbf s}(\sigma)\in{\mathcal S}$, denoting the derivative $\frac{d}{d\sigma}$ by a prime $'$, and define ${\mathbf h}(\sigma)$ as the intersection
of the two-dimensional subspace ${\mathbf s}'^\perp$ of all vectors
orthogonal to the vector  ${\mathbf s}'=\frac{d}{d\sigma}{\mathbf s}(\sigma)$, and ${\mathcal H}$.
This is sensible since, according to (\ref{G1}), ${\mathbf h}(\tau)$ must be orthogonal to $\dot{\mathbf s}(\tau)$
and hence to ${\mathbf s}'$, the tangent to the curve ${\mathcal S}$ being independent
of the parametrization. We assume for simplicity that the intersection of
${\mathbf s}'^\perp$ and ${\mathcal H}$ is unique and leave it to the reader to
consider the more general case of multiple intersections. Hence we have obtained a kind of (local)
synchronization between the two curves ${\mathcal S}$ and ${\mathcal H}$, see Figure \ref{FIGSBR}.
Then we consider the equation of motion (\ref{G9}) that necessarily holds and employ the solution (\ref{G3}) of the inverse problem.
This entails
\begin{equation}\label{G11}
  \tilde{\mathbf h}= {\mathbf s}\times {\mathbf s}'+\lambda(s) {\mathbf s}\stackrel{!}{=}\frac{d\tau}{d\sigma}{\mathbf h}
  \;.
\end{equation}
Taking scalar products of (\ref{G11}) with ${\mathbf s}(\sigma)$  and ${\mathbf h}(\sigma)$ yields $\frac{d\tau}{d\sigma}$ uniquely as
\begin{equation}\label{G12}
  \frac{d\tau}{d\sigma}=\frac{({\mathbf s}\times{\mathbf s}')\cdot {\mathbf h}}{{\mathbf h}\cdot{\mathbf h}-\left({\mathbf s}\cdot{\mathbf h}\right)^2}
  \;,
\end{equation}
neglecting special cases.
Finally, $\tau(\sigma)$ is obtained as the integral $\tau(\sigma)=\int \frac{d\tau}{d\sigma}\,d\sigma$.\\

The general solution of (\ref{G1}) can be obtained as a linear superposition of three ``fundamental solutions"
${\mathbf s}^{(i)}(\tau),\;i=1,2,3$. We will slightly generalize the initial point of time to $\tau=\tau_0$.
Since scalar products are invariant under time evolution according to (\ref{G1})
it follows that the fundamental solutions will form a right-handed orthonormal frame for all $\tau\in{\mathbbm R}$, i.~e.,
\begin{equation}\label{G14}
  R(\tau,\tau_0)\equiv \left( {\mathbf s}^{(1)}(\tau), {\mathbf s}^{(2)}(\tau), {\mathbf s}^{(3)}(\tau)\right)
 \end{equation}
will be a rotational matrix with unit determinant, in symbols, $R(\tau,\tau_0)\in SO(3)$. The three equations of motion (\ref{G1})
for the ${\mathbf s}^{(i)}(\tau),\;i=1,2,3$ can be compactly written in matrix form as
\begin{equation}\label{G15}
  \frac{d}{d\tau}\,R(\tau,\tau_0)=H(\tau)\,R(\tau,\tau_0)
  \;,
\end{equation}
where $H(\tau)$ is the real, antisymmetric $3\times 3$-matrix
\begin{equation}\label{G16}
 H(\tau)=\left(
 \begin{array}{ccc}
   0 & -h_3(\tau) & h_2(\tau) \\
   h_3(\tau) & 0 & -h_1(\tau) \\
   -h_2(\tau) & h_1(\tau) &0
 \end{array}
  \right)
 \;,
\end{equation}
and the $h_i(\tau)$ are the components of the magnetic field according to (\ref{G1a}).
Moreover, we choose the initial values ${\mathbf s}^{(i)}(\tau_0)$ as the corresponding unit standard vectors such that
\begin{equation}\label{G13}
  R(\tau_0,\tau_0)\equiv \left( {\mathbf s}^{(1)}(\tau_0), {\mathbf s}^{(2)}(\tau_0), {\mathbf s}^{(3)}(\tau_0)\right)={\mathbbm 1}
  \;.
 \end{equation}

The general solution ${\mathbf s}(\tau)$ of (\ref{G1}) with initial value ${\mathbf s}(\tau_0)$ can be obtained
from the three fundamental solutions by means of ${\mathbf s}(\tau)=R(\tau,\tau_0)\,{\mathbf s}(\tau_0)$.
More generally, let $t\mapsto\tilde{R}(\tau,\tau_0)\in SO(3)$ be a one-parameter family of rotational matrices
satisfying the differential equation
\begin{equation}\label{G17}
\frac{d}{d\tau}\,\tilde{R}(\tau,\tau_0)=H(\tau)\,\tilde{R}(\tau,\tau_0)
\end{equation}
with initial condition
\begin{equation}\label{G18}
 \tilde{R}(\tau_0,\tau_0)=S\in SO(3)
 \;,
\end{equation}
then it follows that
\begin{equation}\label{G19}
 \tilde{R}(\tau,\tau_0)=R(\tau,\tau_0)\,S
\end{equation}
for all $\tau\in{\mathbbm R}$ since both sides of (\ref{G19}) satisfy the same differential equation
with the same initial value.\\

For the sake of reference we recapitulate the following well-known method of transforming (\ref{G17}) into a rotating frame.
Set $\tau_0=0$ and  let $U(\tau)\in SO(3)$ be such that
\begin{equation}\label{G20}
 \dot{U}=K\,U,
\end{equation}
where $K(\tau)$ is an anti-symmetric $3\times 3$-matrix, moreover
\begin{equation}\label{G21}
 \tilde{R}\equiv U\,R
 \;.
\end{equation}
$\tilde{R}$ can be thought to consist of the columns of $R$ transformed to a frame rotating with $U^\top$.
We obtain for the transformed equation of motion
\begin{equation}\label{G22}
  \frac{d}{d\tau}\tilde{R}=\dot{U}\,R+U\,\dot{R}\stackrel{(\ref{G15})(\ref{G20})}{=}K\,U\,R+U\,H\,R=\left( K+U\,H\,U^\top\right)\tilde{R}
  \equiv \tilde{H}\,\tilde{R}\;,
\end{equation}
where $\tilde{H}$ will be again an anti-symmetric $3\times 3$-matrix representing the transformed magnetic field.

\section{Main part}\label{sec:M}

For the remainder of the paper we consider the case where the three components $h_i(\tau),\;i=1,2,3$ of the magnetic field ${\mathbf h}(\tau)$
are $T\equiv\frac{2\pi}{\omega}$-periodic functions of $\tau$.
For example, periodicity holds if the magnetic field is generated by a plane electro-magnetic wave.
It does {\it not} follow that all solutions ${\mathbf s}(\tau)$ of the
equation of motion (\ref{G1}) will also be $T$-periodic. However, for suitable initial conditions, at least one periodic solution
${\mathbf s}(\tau)$ always exists. Note that, in general, this does not hold for the underlying $s=1/2$ Schr\"odinger equation.

\subsection{Periodic solutions}\label{sec:P}

\subsubsection{Proof and further remarks}\label{proof1}

In order to prove the existence of a periodic solution, we consider the fundamental matrix solution $R(\tau)$ according to (\ref{G14}) with
initial value $R(0)={\mathbbm 1}$. Its value after one period will be the rotational matrix $R(T)\in SO(3)$.
As any rotational matrix with unit determinant it will be a rotation about an axis with an angle $\delta$.
Accordingly, $R(T)$ will have the eigenvalues $1, e^{\pm {\sf i} \delta}$, corresponding to a real eigenvector
${\mathbf a}$ and two generally complex eigenvectors, resp.~. The eigenvector ${\mathbf a}$ satisfying
\begin{equation}\label{P0}
 R(T)\,{\mathbf a}={\mathbf a}
\end{equation}
represents the axis of rotation and, after normalization $\|{\mathbf a}\|=1$, will be unique up to a sign.
The angle of rotation is also only defined up to a sign:
If $\delta$ is the angle of rotation about ${\mathbf a}$
then $-\delta$ will be the angle of rotation about $-{\mathbf a}$. Since the trace of $R(T)$ is the sum of its eigenvalues
it follows that
\begin{equation}\label{P1}
 \mbox{Tr}\, R(T)= 1+e^{{\sf i} \delta}+e^{-{\sf i} \delta}=1+2\,\cos \delta
 \;,
\end{equation}
and hence $\pm \delta$ can immediately be obtained from $R(T)$.
In the special case where $R(T)={\mathbbm 1}$ it follows that $\pm\delta=0$ and $R(T)$ has a three-dimensional
degenerate eigenspace.

For later reference we note that $R(T)$ can be written as the matrix exponential
\begin{equation}\label{P1a}
 R(T)= e^{\delta A}
 \;,
\end{equation}
where, analogously to (\ref{G16}),
\begin{equation}\label{P1b}
 A\equiv\left(
 \begin{array}{ccc}
   0 & -a_3 & a_2\\
   a_3& 0 &  -a_1 \\
   -a_2 & a_1 &0
 \end{array}
  \right)
 \;,
\end{equation}
and the $a_i$ are the components of the normalized eigenvector ${\mathbf a}$ according to (\ref{P0}).
Note that the eigenvalues of $A$ are of the form $0,\pm{\sf i}$.

Now we consider the special solution ${\mathbf s}(\tau)$ of (\ref{G1}) with the initial value given by the real eigenvector
of $R(T)$, namely ${\mathbf s}(0)={\mathbf a}$. According to the remark  after (\ref{G16}) in  Section \ref{sec:G} this solution
can be written as ${\mathbf s}(\tau)=R(\tau)\,{\mathbf a}$ for all $\tau\in{\mathbbm R}$. It follows that
\begin{equation}\label{P2}
 {\mathbf s}(T)=R(T)\,{\mathbf a}\stackrel{(\ref{P0})}{=} {\mathbf a} =  {\mathbf s}(0)
 \;,
\end{equation}
and, more generally,
\begin{equation}\label{P3}
 {\mathbf s}(\tau+T)={\mathbf s}(\tau)
 \;,
\end{equation}
for all $\tau\in{\mathbbm R}$. Hence ${\mathbf s}(\tau)$ is the $T$-periodic solution we are looking for.

In order to prove (\ref{P3}) we restore the notation of $\tau=\tau_0$ for the initial point of time
and write, for example, $R(T+\tau_0,\tau_0)$ instead of $R(T)$.
We consider the one-parameter family $\tau\mapsto R(\tau+T,\tau_0)$ of rotational matrices
with initial value $R(\tau_0+T,\tau_0)$.
It satisfies the
differential equation
\begin{equation}\label{P4}
  \frac{d}{d\tau}\,R(\tau+T,\tau_0)\stackrel{(\ref{G15})}{=} H(\tau+T)\,R(\tau+T,\tau_0)= H(\tau)\,R(\tau+T,\tau_0)
  \;,
\end{equation}
where we have used that $H(\tau)$ is $T$-periodic.
Applying (\ref{G19}) and the preceding remarks we conclude
\begin{equation}\label{P5}
 R(\tau+T,\tau_0)= R(\tau,\tau_0)\,R(\tau_0+T,\tau_0)
 \;.
\end{equation}
Then it follows that
\begin{equation}\label{P6}
{\mathbf s}(\tau+T)=R(\tau+T,t_0)\,{\mathbf s}(\tau_0)\stackrel{(\ref{P5})}{=}
R(\tau,\tau_0)\,R(\tau_0+T,\tau_0)\,{\mathbf a}\stackrel{(\ref{P0})}{=}R(\tau,\tau_0)\,{\mathbf a}=R(\tau,\tau_0)\,{\mathbf s}(\tau_0)={\mathbf s}(\tau)
\;,
\end{equation}
thereby completing the proof of (\ref{P3}).

In general, there will only be two $T$-periodic solutions of unit length, namely ${\mathbf s}(\tau)$ and $-{\mathbf s}(\tau)$.
Only in the special case of $R(T+\tau_0,\tau_0)={\mathbbm 1}$ every solution $\tilde{\mathbf s}(\tau)$  will be $T$-periodic since
every unit vector ${\mathbf a}=\tilde{\mathbf s}(0)$
will satisfy (\ref{P0}) and the preceding arguments may be repeated for $\tilde{\mathbf s}(\tau)$.

To underline the latter statement let $S(\tau)\equiv\left( {\mathbf e}^{(1)}(\tau), {\mathbf e}^{(2)}(\tau), {\mathbf e}^{(3)}(\tau)\right)$ be a solution of
$\dot{S}(\tau)=H(\tau)\,S(\tau)$ such that $S(0)$ is a right-handed orthonormal frame and ${\mathbf e}^{(1)}(\tau)={\mathbf s}(\tau)$
for all $\tau\in{\mathbbm R}$, i.~e., ${\mathbf e}^{(1)}(\tau)$ is the periodic solution considered above. It follows that
$S(T)=\left( {\mathbf a}, {\mathbf e}^{(2)}(T), {\mathbf e}^{(3)}(T)\right)=R(T)\,S(0)$ will also be a right-handed orthonormal frame
and hence ${\mathbf e}^{(2)}(T)$ and ${\mathbf e}^{(3)}(T)$ will lie in the plane ${\mathbf a}^\perp$ orthogonal to ${\mathbf a}$.
But since $R(T)$ is a rotation about the axis through ${\mathbf a}$ with an angle $\delta$, the vectors ${\mathbf e}^{(2)}(T)$ and ${\mathbf e}^{(3)}(T)$
will, in general, be different from  ${\mathbf e}^{(2)}(0)$ and ${\mathbf e}^{(3)}(0)$ resp.~. More precisely,
\begin{eqnarray}
\label{P7a}
 {\mathbf e}^{(2)}(T)&=&\cos(\delta)\,{\mathbf e}^{(2)}(0)+ \sin(\delta)\,{\mathbf e}^{(3)}(0),\\
 \label{P7b}
 {\mathbf e}^{(3)}(T)&=&-\sin(\delta)\,{\mathbf e}^{(2)}(0)+ \cos(\delta)\,{\mathbf e}^{(3)}(0),
\end{eqnarray}
and hence  ${\mathbf e}^{(1)}(\tau)$ and ${\mathbf e}^{(2)}(\tau)$ will not be $T$-periodic except for $\delta=0$.

As a preparation for the next subsection we again consider the fundamental solution $R(\tau,\tau_0)$ satisfying\\
$R(T+\tau_0,\tau_0)=e^{\delta\,A}$ and define
\begin{equation}\label{P8}
  \epsilon\equiv \frac{\delta}{T}
  \;.
\end{equation}
Then we proceed by defining $P(\tau,\tau_0)\in SO(3)$ as
\begin{equation}\label{P9}
P(\tau,\tau_0)\equiv  R(\tau,\tau_0)\,e^{-\epsilon\,(\tau-\tau_0)\,A}
\;,
\end{equation}
and show that it will be $T$-periodic:
\begin{eqnarray}\label{P10a}
 P(\tau+T,\tau_0)&\stackrel{(\ref{P9})}{=}& R(\tau+T,\tau_0)\,e^{-\epsilon\,T\,A}\,e^{-\epsilon\,(\tau-\tau_0)\,A}\stackrel{(\ref{P8})}{=}
 R(\tau+T,\tau_0)\,e^{-\delta\,A}\,e^{-\epsilon\,(\tau-\tau_0)\,A}\\
 \label{P10b}
& \stackrel{(\ref{P5},\ref{P1a})}{=}&
 R(\tau,\tau_0)\,R(\tau_0+T,\tau_0)\,R(\tau_0+T,\tau_0)^{-1}\,e^{-\epsilon\,(\tau-\tau_0)\,A}=
  R(\tau,\tau_0)\,e^{-\epsilon\,(\tau-\tau_0)\,A}\stackrel{(\ref{P9})}{=}
  P(\tau,\tau_0)
  \;.
\end{eqnarray}

\subsubsection{Example 1}\label{ex1}

As an example we consider the circularly polarized Rabi problem where
\begin{equation}\label{P11}
 {\mathbf h}(\tau)=\left(
\begin{array}{c}
 F \cos (\omega  \tau ) \\
 F \sin (\omega  \tau ) \\
 \omega _0 \\
\end{array}
\right)
\;.
\end{equation}
Let $H(\tau)$ denote the corresponding anti-symmetric matrix according to (\ref{G16}).
We will transform the equation of motion into a rotating frame as described in Section \ref{sec:G}
and set
\begin{equation}\label{P11a}
 U(\tau)=\left(
\begin{array}{ccc}
 \cos (\tau  \omega ) & \sin (\tau  \omega ) & 0 \\
 -\sin (\tau  \omega ) & \cos (\tau  \omega ) & 0 \\
 0 & 0 & 1 \\
\end{array}
\right)\;,
\end{equation}
such that
\begin{equation}\label{P11b}
 \dot{U}=K\,U
 \;,
\end{equation}
where
\begin{equation}\label{P11c}
 K=\left(
\begin{array}{ccc}
 0 & \omega  & 0 \\
 -\omega  & 0 & 0 \\
 0 & 0 & 0 \\
\end{array}
\right)\;.
\end{equation}
The transformed equation of motion corresponding to (\ref{G22}) will be written as
\begin{equation}\label{P11d}
  \frac{d}{d\tau}\tilde{S}=\tilde{H}\,\tilde{S}
  \;,
\end{equation}
and $\tilde{H}$ is obtained after a short calculation as
\begin{equation}\label{P11e}
 \tilde{H}=\left(
\begin{array}{ccc}
 0 & \omega -\omega _0 & 0 \\
 \omega _0-\omega  & 0 & -F \\
 0 & F & 0 \\
\end{array}
\right)\;,
\end{equation}
no longer depending on $\tau$. Hence (\ref{P11d}) can be immediately solved
with the result
\begin{equation}\label{P11f}
  \tilde{S}=\left(
\begin{array}{ccc}
 \frac{F}{\Omega } & \left(\frac{\omega }{\Omega }-\frac{\omega _0}{\Omega }\right)
   \sin (\tau  \Omega ) & \left(\frac{\omega }{\Omega }-\frac{\omega _0}{\Omega
   }\right) \cos (\tau  \Omega ) \\
 0 & \cos (\tau  \Omega ) & -\sin (\tau  \Omega ) \\
 \frac{\omega _0-\omega }{\Omega } & \frac{F \sin (\tau  \Omega )}{\Omega } & \frac{F
   \cos (\tau  \Omega )}{\Omega } \\
\end{array}
\right)\;,
\end{equation}
where
\begin{equation}\label{P13}
 \Omega=\Omega_{Rabi}\equiv \sqrt{F^2+\left(\omega_0-\omega\right)^2}
 \;
\end{equation}
denotes the so-called Rabi frequency.
The remaining reverse transformations yield
\begin{equation}\label{P11g}
 S=U^\top\,\tilde{S}
 \;,
\end{equation}
and, in order to obtain a fundamental solution,
\begin{equation}\label{P11h}
 R=S\,S(0)^{-1}
 \;.
\end{equation}
The result is of the somewhat complicated form
\begin{equation}\label{P12}
\footnotesize
R(\tau)=
\left(
\begin{array}{ccc}
 \frac{\cos (\tau  \omega ) \left(F^2+\left(\omega -\omega _0\right){}^2 \cos (\tau
   \Omega )\right)+\left(\omega -\omega _0\right) \Omega  \sin (\tau  \omega ) \sin
   (\tau  \Omega )}{\Omega ^2} & \frac{\sin (\tau  \omega ) \left(F^2+\left(\omega
   -\omega _0\right){}^2 \cos (\tau  \Omega )\right)+\left(\omega _0-\omega \right)
   \Omega  \cos (\tau  \omega ) \sin (\tau  \Omega )}{\Omega ^2} & \frac{F
   \left(\omega -\omega _0\right) (\cos (\tau  \Omega )-1)}{\Omega ^2} \\
 \frac{\left(\omega -\omega _0\right) \cos (\tau  \omega ) \sin (\tau  \Omega
   )}{\Omega }-\sin (\tau  \omega ) \cos (\tau  \Omega ) & \frac{\left(\omega
   -\omega _0\right) \sin (\tau  \omega ) \sin (\tau  \Omega )}{\Omega }+\cos (\tau
   \omega ) \cos (\tau  \Omega ) & \frac{F \sin (\tau  \Omega )}{\Omega } \\
 \frac{F \left(\Omega  \sin (\tau  \omega ) \sin (\tau  \Omega )+2 \left(\omega
   _0-\omega \right) \cos (\tau  \omega ) \sin ^2\left(\frac{\tau  \Omega
   }{2}\right)\right)}{\Omega ^2} & -\frac{F \left(2 \left(\omega -\omega _0\right)
   \sin (\tau  \omega ) \sin ^2\left(\frac{\tau  \Omega }{2}\right)+\Omega  \cos
   (\tau  \omega ) \sin (\tau  \Omega )\right)}{\Omega ^2} & \frac{F^2 \cos (\tau
   \Omega )+\left(\omega -\omega _0\right){}^2}{\Omega ^2} \\
\end{array}
\right).
\end{equation}
\normalsize

After one period the fundamental solution reads
\begin{equation}\label{P14}
R(T)=\left(
\begin{array}{ccc}
 \frac{F^2+\left(\omega -\omega _0\right){}^2 \cos \left(\frac{2 \pi  \Omega
   }{\omega }\right)}{\Omega ^2} & \frac{\left(\omega _0-\omega \right) \sin
   \left(\frac{2 \pi  \Omega }{\omega }\right)}{\Omega } & \frac{2 F \left(\omega
   _0-\omega \right) \sin ^2\left(\frac{\pi  \Omega }{\omega }\right)}{\Omega ^2} \\
 \frac{\left(\omega -\omega _0\right) \sin \left(\frac{2 \pi  \Omega }{\omega
   }\right)}{\Omega } & \cos \left(\frac{2 \pi  \Omega }{\omega }\right) & \frac{F
   \sin \left(\frac{2 \pi  \Omega }{\omega }\right)}{\Omega } \\
 \frac{2 F \left(\omega _0-\omega \right) \sin ^2\left(\frac{\pi  \Omega }{\omega
   }\right)}{\Omega ^2} & -\frac{F \sin \left(\frac{2 \pi  \Omega }{\omega
   }\right)}{\Omega } & \frac{F^2 \cos \left(\frac{2 \pi  \Omega }{\omega
   }\right)+\left(\omega -\omega _0\right){}^2}{\Omega ^2} \\
\end{array}
\right).
\end{equation}
This matrix has the eigenvalues $1,e^{\pm{\sf i}\delta}$
with
\begin{equation}\label{P15}
 \delta=\frac{2\pi\Omega}{\omega}
 \;,
\end{equation}
and the normalized eigenvector
\begin{equation}\label{P16}
 {\mathbf a}=\left(
\begin{array}{c}
 \frac{F}{\Omega } \\
 0 \\
 \frac{\omega _0-\omega }{\Omega } \\
\end{array}
\right)
\end{equation}
corresponding to the eigenvalue $1$. This implies the well-known result for the quasienergy
\begin{equation}\label{P17}
  \epsilon = \frac{\delta}{T}=\Omega \mod \omega
  \;.
\end{equation}
We consider the right-handed orthonormal system with ${\mathbf a}$ as its first vector and write
this system as the columns of a rotational matrix
\begin{equation}\label{P18}
  A=\frac{1}{\Omega}
  \left(
\begin{array}{ccc}
 F & 0 & \omega -\omega _0 \\
 0 & \Omega  & 0 \\
 \omega _0-\omega  & 0 & F \\
\end{array}
\right)
\;.
\end{equation}
For $A$ as the initial value of the differential equation (\ref{G15}) the solution reads,
according to the argument in connection with (\ref{G17}), (\ref{G18}), (\ref{G19}):
\begin{eqnarray}\label{P19a}
  S(\tau)&=&R(\tau)\,A\\
\nonumber
  &=&
  \frac{1}{\Omega}
  \left(
\begin{array}{ccc}
 F \cos (\tau  \omega ) & \left(\omega -\omega _0\right) \cos (\tau  \omega ) \sin
   (\tau  \Omega )-\Omega  \sin (\tau  \omega ) \cos (\tau  \Omega ) & \Omega  \sin
   (\tau  \omega ) \sin (\tau  \Omega )+\left(\omega -\omega _0\right) \cos (\tau
   \omega ) \cos (\tau  \Omega ) \\
 F \sin (\tau  \omega ) & \left(\omega -\omega _0\right) \sin (\tau  \omega ) \sin
   (\tau  \Omega )+\Omega  \cos (\tau  \omega ) \cos (\tau  \Omega ) & \left(\omega
   -\omega _0\right) \sin (\tau  \omega ) \cos (\tau  \Omega )-\Omega  \cos (\tau
   \omega ) \sin (\tau  \Omega ) \\
 \omega _0-\omega  & F \sin (\tau  \Omega ) & F \cos (\tau  \Omega ) \\
\end{array}
\right) \\
\label{P19b}
 &&
\;.
\end{eqnarray}
Its first column represent the $T$-periodic solution of (\ref{G15}), unique up to a sign,
whereas the second and third columns are not periodic but after one period will advance by the angle
$\delta=T\,\Omega$.

\subsection{Floquet theory}\label{sec:F}

It will be instructive to rephrase the results of the last subsection \ref{sec:P} in terms of Floquet theory, see
\cite{Floquet83} or \cite{YakubovichStarzhinskii75} for a more recent reference.
Floquet theory deals with linear differential equations of the form (\ref{G15}) with a $T$-periodic matrix function $H(\tau)$
but for arbitrary finite dimensions and real or complex matrix functions $R(\tau,\tau_0)$. Its main result is that the fundamental solution
of the differential equation can be written as the product of a $T$-periodic matrix function and a special exponential matrix
function of $t$. In our case of the differential equation (\ref{G15}) this means that the fundamental solution $R(\tau,\tau_0)$
with initial condition $R(\tau_0,\tau_0)={\mathbbm 1}$ can be cast in the ``Floquet normal form"
\begin{equation}\label{F1}
 R(\tau,\tau_0)=P(\tau,\tau_0)\,e^{\epsilon\,(\tau-\tau_0)\,A}
 \;,
\end{equation}
such that $\tau\mapsto P(\tau,\tau_0)\in SO(3)$ is $T$-periodic,  $\epsilon$ is a real number and
$A$ a real anti-symmetric $3\times 3$-matrix.

$\epsilon$ is called a ``Floquet exponent" or, in physical applications, also a ``quasienergy".
For $\tau=\tau_0$ we have ${\mathbbm 1}=R(\tau_0,\tau_0)=P(\tau_0,\tau_0)$.
For $\tau-\tau_0=T$ it follows that
\begin{equation}\label{F4}
 R(\tau_0+T,\tau_0)\stackrel{(\ref{F1})}{=} P(\tau_0+T,\tau_0)\,e^{\epsilon\,T\,A}=P(\tau_0,\tau_0)\,e^{\epsilon\,T\,A}\stackrel{(\ref{P1a})}{=}
 {\mathbbm 1}\,e^{\delta\,A}=e^{\delta\,A}
 \;.
\end{equation}

Recall that we have defined $A, \delta$ and $P(\tau,\tau_0)$ in accordance with these requirements in subsection \ref{sec:P},
and thus we have proven the main result of Floquet theory for the special case of the classical Rabi problem.

The quasienergy $\epsilon$ is only defined up to integer multiples of $\omega$ because, for $n\in{\mathbbm Z}$,
\begin{equation}\label{F7}
 R(\tau,\tau_0)=\left( P(\tau,\tau_0)\,e^{- n\,\omega\,(\tau-\tau_0)\,A}\right)\,\left(e^{(\epsilon+ n\,\omega)\,(\tau-\tau_0)\,A}\right)
 \end{equation}
would also be a product of a periodic and an exponential matrix function and hence also of the Floquet normal form.
Here we have used the fact that the eigenvalues of $A$ are $0,\pm {\sf i}$.
Similarly, the replacement ${\mathbf a}\mapsto -{\mathbf a}$ and hence $A\mapsto -A$  shows that together with $\epsilon$,
$-\epsilon$ is also a possible quasienergy. Taking into account the non-uniqueness of the quasienergy we also consider the
equivalence class
\begin{equation}\label{F8}
  [\epsilon]\equiv \{\epsilon+n\,\omega\left| n\in{\mathbbm Z}\right.\}
  \;.
\end{equation}

In applications to the Schr\"dinger equation the operator $e^{\epsilon\,(\tau-\tau_0)\,A}$ is often
decomposed in terms of its eigenvalues and eigenvectors. In the context of the classical Rabi problem we
will only consider its real eigenvector ${\mathbf a}$ corresponding to the eigenvalue $e^{\epsilon\,(\tau-\tau_0)}$
and not the other two complex eigenvectors and eigenvalues since they have no direct geometric interpretation.

\subsection{Quasienergy I}\label{sec:QI}

It has been shown \cite{S18} that the quasienergy of the  $s=1/2$ Schr\"odinger equation
with a periodic magnetic field can be expressed in terms of integrals using the periodic solution of the analogous classical Rabi problem.
Here we will re-derive the analogous result for the classical Rabi problem
without employing the reference to the Schr\"odinger equation, solely by using the periodic solution
${\mathbf s}(\tau)$ considered in subsection \ref{sec:P}.

To this end we consider the time-dependent right-handed orthonormal frame, shortly called ``${\mathbf e}$-frame", defined by
\begin{eqnarray}
\label{QI1a}
{\mathbf e}^{(1)}(\tau) &=& {\mathbf s}(\tau), \\
  \label{QI1b}
{\mathbf e}^{(2)}(\tau) &=& \frac{\dot{\mathbf s}(\tau)}{\|\dot{\mathbf s}(\tau)\|} ,\\
     \label{QI1c}
  {\mathbf e}^{(3)}(\tau) &=& {\mathbf e}^{(1)}(\tau)\times {\mathbf e}^{(2)}(\tau)
  \;.
\end{eqnarray}
Further, let
\begin{equation}\label{QI2}
  S(\tau)=\left( {\mathbf s}^{(1)}(\tau), {\mathbf s}^{(2)}(\tau), {\mathbf s}^{(3)}(\tau)\right)\in SO(3)
\end{equation}
be a solution of (\ref{G15}) with initial conditions
\begin{equation}\label{QI3}
   {\mathbf s}^{(i)}(0)={\mathbf e}^{(i)}(0),\quad i=1,2,3
   \;,
\end{equation}
and hence
\begin{equation}\label{QI4}
  {\mathbf s}^{(1)}(\tau)={\mathbf e}^{(1)}(\tau)={\mathbf s}(\tau)
\end{equation}
for all $\tau\in{\mathbbm R}$. It follows that the other two components of $S(\tau)$ can be expanded
w.~r.~t.~the  ${\mathbf e}$-frame in the form
\begin{eqnarray}
\label{QI5a}
{\mathbf s}^{(2)}(\tau) &=& \cos(\alpha(\tau))\,{\mathbf e}^{(2)}(\tau)+\sin(\alpha(\tau))\,{\mathbf e}^{(3)}(\tau),\\
   \label{QI5b}
{\mathbf s}^{(3)}(\tau) &=& -\sin(\alpha(\tau))\,{\mathbf e}^{(2)}(\tau)+\cos(\alpha(\tau))\,{\mathbf e}^{(3)}(\tau)
\;,
\end{eqnarray}
where $\alpha(\tau)$ is a smooth function given by an integral that we will derive below.
We use the abbreviation
\begin{equation}\label{QI6}
v(\tau)^2\equiv \dot{\mathbf s}(\tau)\cdot \dot{\mathbf s}(\tau)\quad\Rightarrow\quad v\,\dot{v}= \dot{\mathbf s}\cdot \ddot{\mathbf s}
\;,
\end{equation}
and expand the magnetic field w.~r.~t.~the ${\mathbf e}$-frame:
\begin{equation}\label{QI7}
{\mathbf h}(\tau)=\sum_{i=1}^{3} k_i(\tau) {\mathbf e}^{(i)}(\tau)
\;.
\end{equation}
The equation
\begin{equation}\label{QI8}
 \dot{\mathbf s}= v\,{\mathbf e}^{(2)}={\mathbf h}\times {\mathbf s}=\left( k_1 {\mathbf e}^{(1)}+ k_2 {\mathbf e}^{(2)}+ k_3 {\mathbf e}^{(3)}\right)
 \times  {\mathbf e}^{(1)}= -k_2 {\mathbf e}^{(3)}+k_3 {\mathbf e}^{(2)}
 \end{equation}
immediately implies
\begin{equation}\label{QI9}
 k_1={\mathbf h}\cdot{\mathbf s},\quad k_2=0,\quad k_3=v
 \;.
\end{equation}
We will also expand $\ddot{\mathbf s}$ w.~r.~t.~the ${\mathbf e}$-frame. First, we obtain
\begin{equation}\label{QI10}
 \dot{\mathbf s}\cdot {\mathbf s}=0\;\Rightarrow\;\dot{\mathbf s}\cdot \dot{\mathbf s}+\ddot{\mathbf s}\cdot {\mathbf s}=0
 \;\Rightarrow\;\ddot{\mathbf s}\cdot {\mathbf s}=-v^2
 \;.
\end{equation}
Second,
\begin{equation}\label{QI11}
 \ddot{\mathbf s}\cdot {\mathbf e}^{(3)}=\ddot{\mathbf s}\cdot \left({\mathbf e}^{(1)} \times {\mathbf e}^{(2)} \right)
 =\ddot{\mathbf s}\cdot \left({\mathbf s} \times \frac{1}{v}\dot{\mathbf s}\right)
 =\frac{1}{v} {\mathbf s}\cdot\left( \dot{\mathbf s}\times \ddot{\mathbf s}\right) \equiv \frac{g}{v}
 \;,
\end{equation}
where we have abbreviated the triple product $ {\mathbf s}\cdot\left( \dot{\mathbf s}\times \ddot{\mathbf s}\right)$
by $g$. Together with (\ref{QI6}) the last two equations yield
\begin{equation}\label{QI12}
\ddot{\mathbf s}=-v^2\,{\mathbf e}^{(1)}+\dot{v}\,{\mathbf e}^{(2)}+ \frac{g}{v}\, {\mathbf e}^{(3)}
\;.
\end{equation}
With this we can evaluate the $\tau$-derivatives of the frame vectors in the following way:
\begin{equation}\label{QI13}
  \dot{\mathbf e}^{(2)}=\frac{d}{d\tau}\frac{\dot{\mathbf s}}{v}=-\frac{\dot{v}}{v^2}\dot{\mathbf s}+\frac{1}{v}\ddot{\mathbf s}
  \stackrel{(\ref{QI12})}{=}-v\,{\mathbf e}^{(1)}+\frac{g}{v^2}\,{\mathbf e}^{(3)}
  \;,
\end{equation}

\begin{equation}\label{QI14}
  \dot{\mathbf e}^{(3)}\stackrel{(\ref{QI1c})}{=}\dot{\mathbf e}^{(1)}\times {\mathbf e}^{(2)}+{\mathbf e}^{(1)}\times \dot{\mathbf e}^{(2)}
  =\frac{1}{v}\underbrace{\dot{\mathbf s}\times \dot{\mathbf s}}_{0}+\frac{g}{v^2}\,{\mathbf e}^{(1)}\times {\mathbf e}^{(3)}
  =-\frac{g}{v^2}\,{\mathbf e}^{(2)}
  \;.
\end{equation}

The $\tau$-derivative of ${\mathbf s}^{(2)}(\tau)$ can now be calculated in two different ways:
\begin{eqnarray}
\label{QI15a}
  \dot{\mathbf s}^{(2)} &=& {\mathbf h}\times {\mathbf s}^{(2)} \\
  \label{QI15b}
   &\stackrel{(\ref{QI9},\ref{QI5a})}{=}& \left( ({\mathbf h}\cdot{\mathbf s})\,{\mathbf e}^{(1)}+v\, {\mathbf e}^{(3)}\right)
   \times\left( \cos\alpha\, {\mathbf e}^{(2)}+\sin\alpha\,{\mathbf e}^{(3)}\right) \\
   \label{QI15c}
  &=& - v\,\cos\alpha\,{\mathbf e}^{(1)}-({\mathbf h}\cdot{\mathbf s})\,\sin\alpha \, {\mathbf e}^{(2)}+
  ({\mathbf h}\cdot{\mathbf s})\,\sin\alpha \, {\mathbf e}^{(3)}
  \;,
\end{eqnarray}
and
\begin{eqnarray}
\label{QI16a}
  \dot{\mathbf s}^{(2)} &\stackrel{(\ref{QI5a})}{=}&-\dot{\alpha}\,\sin\alpha\,{\mathbf e}^{(2)}+\cos\alpha\,\dot{\mathbf e}^{(2)}
  +\dot{\alpha}\,\cos\alpha\,{\mathbf e}^{(3)}+\sin\alpha\,\dot{\mathbf e}^{(3)}\\
  \label{QI16b}
  &\stackrel{(\ref{QI13},\ref{QI14})}{=}&
  \dot{\alpha}\left(-\sin\alpha\,{\mathbf e}^{(2)} +\cos\alpha\,{\mathbf e}^{(3)}\right)+
  \cos\alpha\left(-v \, {\mathbf e}^{(1)}+\frac{g}{v^2}\,{\mathbf e}^{(3)} \right)
  -\sin\alpha\,\frac{g}{v^2}\,{\mathbf e}^{(2)}
  \;.
\end{eqnarray}
Comparing the ${\mathbf e}^{(2)}$-components of (\ref{QI15c}) and (\ref{QI16b}) yields the expression for $\dot{\alpha}$ we are looking for:
\begin{equation}\label{QI14}
  \dot{\alpha}=({\mathbf h}\cdot{\mathbf s})-\frac{g}{v^2}
  =({\mathbf h}\cdot{\mathbf s})-\frac{ {\mathbf s}\cdot\left( \dot{\mathbf s}\times \ddot{\mathbf s}\right) }{\dot{\mathbf s}\cdot\dot{\mathbf s}}
  \;.
\end{equation}
This shows that $\alpha(\tau)$ can be obtained as an integral over the r.~h.~s.~of (\ref{QI14}) that is a function of ${\mathbf s}(\tau)$ and its
first two derivatives. Note that we have not used the fact that ${\mathbf s}(\tau)$ would be $T$-periodic. These calculations show that if one solution
${\mathbf s}(\tau)$ of (\ref{G1}) is given, then the other two solutions with orthogonal initial conditions can be obtained by means of certain integrals.

In particular we obtain the quasienergy as
\begin{equation}\label{QI15}
  \epsilon= \frac{\delta}{T}=\frac{1}{T}\int_{0}^{T}
  \left(({\mathbf h}\cdot{\mathbf s})-\frac{ {\mathbf s}\cdot\left( \dot{\mathbf s}\times \ddot{\mathbf s}\right) }{\dot{\mathbf s}\cdot\dot{\mathbf s}}
  \right)\,d\tau
  \;,
\end{equation}
now assuming that ${\mathbf s}(\tau)$ will be $T$-periodic.
This result slightly improves the corresponding equation (62) in \cite{S18} in so far as it is manifestly independent of a coordinate system.
The explicit accordance with \cite{S18} will be shown later in subsection \ref{sec:QII}.
We note that the form of (\ref{QI15})  suggest the following splitting of the quasienergy
\begin{equation}\label{QI16}
  \epsilon= \epsilon_d+\epsilon_g\equiv
 \frac{1}{T}\int_{0}^{T}
{\mathbf h}\cdot{\mathbf s}\,d\tau
  +
  \frac{1}{T}\int_{0}^{T}
  \left(-\frac{ {\mathbf s}\cdot\left( \dot{\mathbf s}\times \ddot{\mathbf s}\right) }{\dot{\mathbf s}\cdot\dot{\mathbf s}}
  \right)\,d\tau
\end{equation}
into a ``dynamical part" $\epsilon_d$ and a ``geometrical part" $\epsilon_g$. The dynamical part $\epsilon_d$ is obviously
the time average of the energy ${\mathbf h}\cdot{\mathbf s}$, and depends on the dynamics, i.~e., how fast the angle between
${\mathbf s}$ and ${\mathbf h}$ changes over one period.

For the geometrical part $\epsilon_d$ we note that the
corresponding integral $\int_{0}^{T}\ldots d\tau$ is invariant under an arbitrary parameter transformation $\tau\mapsto \sigma(\tau)$
that leads to a new period $\tilde{T}$:
\begin{equation}\label{QI17}
  \int_{0}^{T}
  \left(-\frac{ {\mathbf s}\cdot\left( \dot{\mathbf s}\times \ddot{\mathbf s}\right) }{\dot{\mathbf s}\cdot\dot{\mathbf s}}
  \right)\,d\tau=
   \int_{0}^{\tilde{T}}
  \left(-\frac{ {\mathbf s}\cdot\left( {\mathbf s}'\times {\mathbf s}''\right) }{{\mathbf s}'\cdot{\mathbf s}'}
  \right)\,d\sigma
  \;,
\end{equation}
where we have denoted the $\sigma$-derivative by a prime $'$. This transformation produces a factor
$\left(\frac{d\sigma}{d\tau}\right)^3$ in the numerator of the integrand of the l.~h.~s.~of (\ref{QI17})
and a factor $\left(\frac{d\sigma}{d\tau}\right)^2$ in the denominator;
after cancelling the remaining factor $\frac{d\sigma}{d\tau}$ is used to transform the $d\tau$-integration
into a $d\sigma$-integration.
This means that this integral is independent of the dynamics
of the spin precession and depends only on the geometry of the curve ${\mathcal S}$,
thereby justifying the denotation as ``geometrical part of the quasienergy".
Note, however, that $\epsilon_g$ still depends on the period $T$ according to the pre-factor $\frac{1}{T}$ in (\ref{QI16}).
The geometrical meaning of $\epsilon_g$ will be considered below.

Next we remark that the splitting $\epsilon=\epsilon_d+\epsilon_g$ can be connected to the form of the magnetic field
according to the solution of the inverse problem (\ref{G3}),
\begin{equation}\label{QI18}
 {\mathbf h}(\tau)={\mathbf s}\times\dot{\mathbf s}+\lambda(\tau)\,{\mathbf s}
 \;,
\end{equation}
where now the function $\lambda(\tau)$ must be $T$-periodic.
First we note that due to ${\mathbf s}\cdot\left({\mathbf s}\times \dot{\mathbf s}\right)= 0$
\begin{equation}\label{QI19}
{\mathbf h}\cdot{\mathbf s}=\lambda(\tau){\mathbf s}\cdot{\mathbf s}=\lambda(\tau)
\;,
\end{equation}
and hence the dynamical part $\epsilon_d$
can be obtained by an integration that involves only the second part of ${\mathbf h}(\tau)$ according to (\ref{QI18}).
Recall that for a given spin function ${\mathbf s}(\tau)$  there are different magnetic fields ${\mathbf h}(\tau)$ satisfying
the equation of motion (\ref{G1}), corresponding to different functions $\lambda(\tau)$ in (\ref{QI18}). 
Upon the choice of these magnetic fields one can realize any value of $\epsilon_d\in{\mathbbm R}$.

In contrast, the integral defining the geometrical part $\epsilon_g$ depends only on the curve ${\mathcal S}$,
hence for its calculation we may use any parametrization of ${\mathcal S}$ and
any magnetic field that satisfies the corresponding equation of motion.

The following choice considerably simplifies the geometry: As a parameter of
${\mathcal S}$ we will use the arc length that will always be denoted by $s$ in what follows.
Differentiation w.~r.~t.~$s$ will again be denoted by a prime $'$ without danger of confusion.
The length of the curve ${\mathcal S}$ will be denoted by $\left|{\mathcal S}\right|$.
This has the consequence that
\begin{equation}\label{QI19a}
 {\mathbf s}'(s)=v=1 \quad \mbox{for all } s\in[0,\left|{\mathcal S}\right|)
 \;.
\end{equation}
Further we will choose as a corresponding magnetic field the first
part of (\ref{QI18}), namely
\begin{equation}\label{QI20}
  {\mathbf h}(s)={\mathbf s}(s)\times{\mathbf s}'(s)
  \;,
\end{equation}
which will always be a unit vector,
\begin{equation}\label{QI21}
\| {\mathbf h}(s)\|=1
\;,
\end{equation}
as the vector product (\ref{QI20}) of two orthogonal unit vectors.
Due to ${\mathbf h}\cdot{\mathbf s}=0$ the dynamical part $\epsilon_d$ of the quasienergy always vanishes. 
For reasons that will become clear later we call (\ref{QI20}) the ``dual magnetic field" of the
spin vector function ${\mathbf s}(s)$ and ${\mathcal H}$ the ``dual loop" of the loop ${\mathcal S}$.
With the above choice the geometric part of the quasienergy can be written as
\begin{equation}\label{QI22}
 \epsilon_g=\frac{1}{T}\int_{0}^{\left|{\mathcal S}\right|}\left(-{\mathbf s}\cdot\left({\mathbf s}'\times {\mathbf s}''\right)\right)\,ds
 \;.
\end{equation}

It is known from elementary differential geometry that the ``geodesic curvature" $k_g$ of a curve ${\mathcal S}$ on a surface parametrized by its arc length
is defined as the triple product
\begin{equation}\label{defgc}
 k_g\equiv {\mathbf s}\cdot\left({\mathbf s}'\times {\mathbf s}''\right)
\end{equation}
measuring the component of the acceleration ${\mathbf s}''$ in the tangent plane of the curve, see, e.~g., \cite{MP77}.
It can have positive or negative values and vanishes at the inflection points of the curve.
It follows that the integrand in (\ref{QI22}) can be, up to a sign,
interpreted as the geodesic curvature $k_g$ of ${\mathcal S}$.
Then we will apply the theorem of Gauss-Bonnet \cite{MP77} for the unit sphere that may be written as
\begin{equation}\label{GaussBonnet}
\int_{M} K\,dA +\int_{\partial M}k_g\,ds=2\pi
  \;.
\end{equation}
Here $M$ denotes a two-dimensional submanifold of $S^2$ with boundary $\partial M$ and (constant) Gaussian curvature $K$.
In our case we set $\partial M={\mathcal S}$ and can identify the surface integral
$\int_{M} K\,dA$ with the (signed) solid angle ${\mathcal A}({\mathcal S})$
encircled by the loop ${\mathcal S}$ and thus re-write (\ref{GaussBonnet}) in the form
\begin{equation}\label{GaussBonnetApp}
\int_{0}^{\left|{\mathcal S}\right|}\left(-{\mathbf s}\cdot\left({\mathbf s}'\times {\mathbf s}''\right)\right)
=-\int_{{\mathcal S}}k_g\,ds\stackrel{(\ref{GaussBonnet})}{=}{\mathcal A}({\mathcal S})-2\pi
  \;.
\end{equation}
The last term $-2\pi$ is irrelevant since the quasienergy is only defined modulo $\omega=\frac{2\pi}{T}$.
Thus we thus have re-established the result
\begin{equation}\label{QI23}
 \epsilon_g=\frac{1}{T}{\mathcal A}({\mathcal S}) \mbox{ modulo } \omega
 \;,
\end{equation}
that endows $\epsilon_g$ with a geometric meaning.\\

One may ask whether the full strength of the theorem of Gauss-Bonnet is really necessary to prove (\ref{QI23}).
For its essential content is the validity for \textit{general} surfaces whereas we need only its application to the sphere $S^2$.
We are not aware of a direct precursor of the Gauss-Bonnet theorem restricted to $S^2$. But, as has been pointed out in
\cite{BB07}, analogous results for geodesic polygons in spherical geometry, especially for triangles, have been known long before:
The area ${\mathcal A}(\Delta)$ of a spherical triangle $\Delta$ equals its \textit{spherical excess},
\begin{equation}\label{QI24}
  {\mathcal A}(\Delta)=\alpha_1+\alpha_2+\alpha_3\,-\,\pi
  \;,
\end{equation}
where the $\alpha_i$ are the interior angles of $\Delta$. This theorem is usually ascribed
to the Flemish mathematician Albert Girard ($1595$ -- $1632$). It can be immediately generalized to
spherical $N$-polygons $\Pi$: Since $\Pi$ can be decomposed into $N-2$ triangles and the area as well as the 
sum over the interior angles are additive, we obtain
\begin{equation}\label{QI25}
   {\mathcal A}(\Pi)=\sum_{i=1}^{N}\alpha_i -(N-2)\pi=\sum_{i=1}^{N}\left(\alpha_i-\pi\right)+2\,\pi
   \;.
\end{equation}
From this (\ref{GaussBonnet}) follows if the curve $\partial M$ is arbitrarily close approximated by
$N$-polygons and if 
\begin{equation}\label{QI26}
  \sum_{i=1}^{N}\left(\alpha_i-\pi\right) \rightarrow -\int_{\partial M} k_g ds
  \;.
\end{equation}
We will use this line of reasoning, following \cite{BB07}, to give a simplified account of the geometric phase
associated to the classical Rabi problem in the next section \ref{sec:GP}.

\subsection{Geometric phases}\label{sec:GP}

\begin{figure}[t]
\centering
\includegraphics[width=0.8\linewidth]{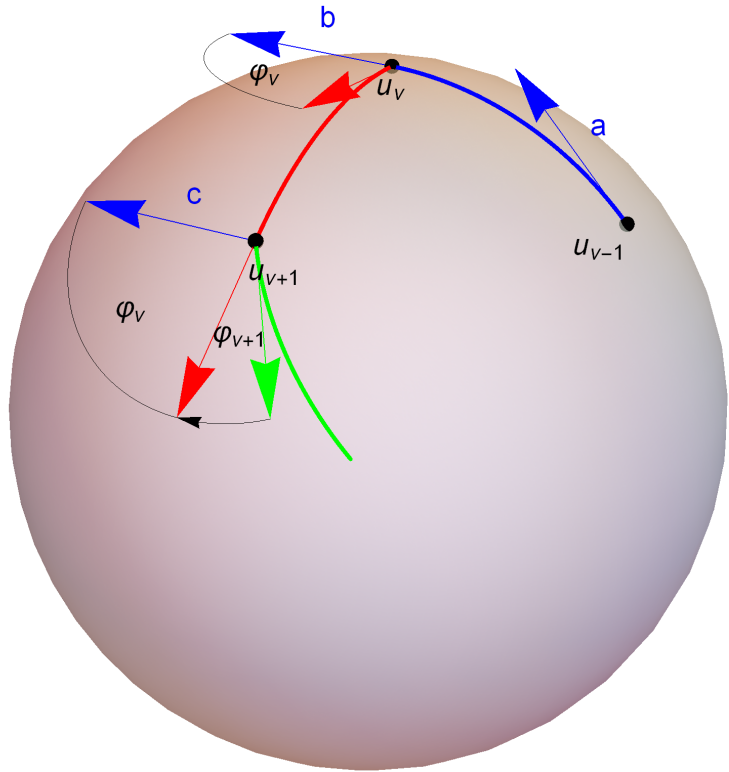}
\caption{Illustration of the parallel transport ${\mathbf a} \to {\mathbf b}\to {\mathbf c}$ along a geodesic polygon on the unit sphere.
}
\label{FIGPT}
\end{figure}

It is well-known \cite{MB14}, \cite{S18} that, in the case of the underlying $s=1/2$ Schr\"odinger equation,
the geometrical part of the quasienergy is closely related to the geometric phase \cite{B84}, \cite{AA87} arising from closed curves
on the unit sphere. Hence it is very plausible that the analogous relation also holds in the case of the classical Rabi problem
thus extending the geometric interpretation given in the last subsection.
It will be instructive to recapitulate this connection without resorting to the Schr\"odinger equation.
Although the pertaining differential-geometrical tools have become somewhat familiar to theoretical physicists due to their
applications in general relativity and gauge theories it is a challenge to give a more elementary account going back to older results
of spherical geometry, partially following \cite{BB07}. Our treatment will not be completely rigorous w.~r.~t.~the standards of pure mathematics,
but this seems to be adequate in view of the purpose.\\

For two points ${\mathbf r}$ and ${\mathbf s}$ on the unit sphere $S^2$ let $T_{\mathbf r}S^2$ and $T_{\mathbf s} S^2$
denote the respective tangent spaces. They are isomorphic as two-dimensional, oriented Euclidean vector spaces but there is
no natural isomorphism $\rho: T_{\mathbf r}S^2\longrightarrow T_{\mathbf s} S^2$. Now consider a smooth curve
${\mathcal P}={\mathcal P}({\mathbf r},{\mathbf s})\subset S^2$ connecting ${\mathbf r}$ and ${\mathbf s}$.
In contrast to the general situation, the curve  ${\mathcal P}$ allows one to define the
\textit{parallel transport} along ${\mathcal P}$ as an isomorphism
\begin{equation}\label{GP1}
 \rho_{\mathcal P}: T_{\mathbf r}S^2\longrightarrow T_{\mathbf s} S^2
 \;.
\end{equation}
Upon introducing arbitrary right-handed, orthonormal basis in $T_{\mathbf r}S^2$ and $T_{\mathbf s} S^2$,
$\rho_{\mathcal P}$ can be described by a rotation with an angle ${\varphi}({\mathcal P})$ that will be called a ``phase shift".
The extension of parallel transport to the case where ${\mathcal P}$  is only piecewise smooth
is in a straightforward way defined by composition of the corresponding transport isomorphisms for smooth curves.

$\rho_{\mathcal P}$ is uniquely
determined by the requirement, additionally to being an isomorphism of two-dimensional, oriented Euclidean vector spaces,
that it maps unit tangent vectors onto unit tangent vectors if $\mathcal P$ is a (segment of a) great circle of $S^2$.
More precisely, let ${\mathbf p}(s),\;s\in[s_1,s_2]$ be the parametrization of a segment of a great circle by the arc length $s$
such that
\begin{equation}\label{GP2}
{\mathbf p}'(s_1)\in T_{{\mathbf p}(s_1)}S^2,\quad \mbox{and  }{\mathbf p}'(s_2)\in T_{{\mathbf p}(s_2)}S^2,
\end{equation}
are the corresponding tangent vectors, necessarily satisfying
\begin{equation}\label{GP3}
  \|{\mathbf p}'(s_1)\|= \|{\mathbf p}'(s_2)\|=1
  \;,
\end{equation}
then we require
\begin{equation}\label{GP4}
  \rho_{\mathcal P}\left({\mathbf p}'(s_1)\right)= {\mathbf p}'(s_2)
  \;.
\end{equation}
It is clear that (\ref{GP4}) uniquely determines the parallel transport along great circles. In fact,
if ${\mathbf v}\in T_{{\mathbf p}(s_1)}S^2$ is a vector that is obtained from ${\mathbf p}'(s_1)$
by a clock-wise rotation with an angle $\gamma$ then $\rho_{\mathcal P}({\mathbf v})$ must be
obtained from ${\mathbf p}'(s_2)$ in the same way.

Next consider a general smooth curve ${\mathcal P}$. It can be approximated arbitrarily close by a polygon $\Pi\in S^2$
consisting piecewise of segments of great circles $\Pi_\nu,\;\nu=1,\ldots,N$.
Recall that great circles are exactly the geodesics of $S^2$ equipped with the natural Riemannian metric.
As mentioned above,
$\rho_\Pi$ will be uniquely defined as the composition
\begin{equation}\label{GP5}
  \rho_\Pi=\rho_{\Pi_N}\circ\ldots\circ \rho_{\Pi_2}\circ\rho_{\Pi_1}
  \;.
\end{equation}
The parallel transport $\rho_{\mathcal P}: T_{\mathbf r}S^2\longrightarrow T_{\mathbf s} S^2$ will
then be defined as the limit of the $\rho_\Pi$ for $N\longrightarrow\infty$ and is such uniquely determined.

We will have a closer look at the geometry of the parallel transport along polygons. Let
${\mathbf u}_\nu$ be the point of intersection between $\Pi_\nu$ and $\Pi_{\nu+1}$
and $d_\nu\equiv d({\mathbf u}_\nu,{\mathbf u}_{\nu+1})$ the spherical distance of ${\mathbf u}_\nu$ and ${\mathbf u}_{\nu+1}$,
i.~e., the arc length of $\Pi_{\nu+1}$ between ${\mathbf u}_\nu$ and ${\mathbf u}_{\nu+1}$.
Further let $\varphi_\nu$ be
the angle between the respective tangent vectors at ${\mathbf u}_\nu$, see Figure \ref{FIGPT}. Then the
tangent vector ${\mathbf a}\in T_{u_{\nu-1}}S^2$ will be parallel transported into
${\mathbf b}\in T_{u_{\nu}}S^2$ and further into ${\mathbf c}\in T_{u_{\nu+1}}S^2$, see Figure \ref{FIGPT}.
We see that the angles between the transported vector and the respective tangent vectors add up as partial sums
of the $\varphi_\nu$.

Now consider a closed polygon $\Pi$, set ${\mathbf u}_0={\mathbf u}_N\equiv {\mathbf u}$ and
consider the parallel transport isomorphism
$\rho_\Pi:T_{\mathbf u}S^2\longrightarrow T_{\mathbf u} S^2$. The total phase shift $\varphi(\Pi)$
will be independent of any local frames and is obtained as
\begin{equation}\label{GP6}
\varphi(\Pi)=\sum_{\nu=1}^{N}\varphi_\nu\;.
\end{equation}
This phase shift can be identified with the ``total geodesic curvature" of the closed polygon $\Pi$.
It vanishes for a full great circle.

A mentioned above, it is at least plausible that for $N\rightarrow\infty$ the quotient $\frac{\varphi_\nu}{d_ \nu}$
approaches a limit which can  be identified with the negative geodesic curvature $k_g$ introduced above
\begin{equation}\label{GP7}
 \frac{\varphi_\nu}{d_ \nu} \rightarrow - k_g(r)
 \;.
\end{equation}
The negative sign in (\ref{GP7}) is due to the that a positively oriented polygon has negative phase shifts $\varphi_\nu$,
see Figure \ref{FIGPT}.
Then the r.~h.~s.~of
\begin{equation}\label{GP8}
\varphi(\Pi)=\sum_{\nu=1}^{N}\varphi_\nu \rightarrow -\int_{0}^{\left|{\mathcal P}\right|}k_g(s)\, ds=\delta({\mathcal P})
\end{equation}
can be considered as the total geodesic curvature of ${\mathcal P}$ that also appears in (\ref{GaussBonnet}).\\

Next we will translate and extend these considerations into the realm of the classical Rabi problem.
Let, as in subsection \ref{sec:QI}, ${\mathcal S}\subset S^2$ be a loop of spin vectors parametrized
as ${\mathbf s}(s)$ by the arc length $s$ for $s\in[0,\left|{\mathcal S}\right|)$ and
${\mathbf h}(s)={\mathbf s}(s)\times{\mathbf s}'(s)$ its dual magnetic field such that
\begin{equation}\label{GP9}
 {\mathbf s}'(s)={\mathbf h}(s)\times{\mathbf s}(s)
 \;.
\end{equation}
The loop of magnetic field vectors ${\mathbf h}(s)$ will again be denoted by ${\mathcal H}$.
We will make it plausible that the parallel transport along the curve ${\mathcal S}$ can be
realized by a solution of the classical Rabi problem using the dual magnetic field.
As above, this will be achieved by approximating ${\mathcal S}$ by a geodesic polygon.

To this end we will divide the parameter range $[0,\left|{\mathcal S}\right|)$ into $N$ equal parts of length
$\delta s=\frac{\left|{\mathcal S}\right|}{N}$ and define
\begin{equation}\label{GP10}
 {\mathbf s}_\nu\equiv {\mathbf s}(s_\nu)\equiv {\mathbf s}(\nu\,\delta s),\quad  {\mathbf s}'_\nu\equiv {\mathbf s}'(s_\nu)
 \;,
\end{equation}
for $\nu=1,\ldots,N$. We will approximate the loop ${\mathcal S}$ by the closed polygon $\Pi$
consisting of segments of great circles $\Pi_\nu$ that are tangent to ${\mathcal S}$ at ${\mathbf s}_\nu$
and oriented into the direction ${\mathbf s}'_\nu$.  As usual, the {\em normal} ${\mathbf n}_\nu$  of  $\Pi_\nu$ is defined as
\begin{equation}\label{GP10a}
 {\mathbf n}_\nu\equiv {\mathbf s}_\nu\times{\mathbf s}_\nu'\stackrel{(\ref{GP9})}{=}
  {\mathbf s}_\nu\times\left({\mathbf h}_\nu\times{\mathbf s}_\nu\right)={\mathbf h}_\nu-
  \underbrace{{\mathbf s}_\nu\cdot{\mathbf h}_\nu}_0\,{\mathbf s}_\nu={\mathbf h}_\nu
  \;.
\end{equation}
It follows  that the normal can be identified with the dual magnetic field vector,
\begin{equation}\label{GP11}
  {\mathbf n}_\nu={\mathbf h}_\nu\equiv {\mathbf h}(r_\nu)
\end{equation}
for all $\nu=1,\ldots, N$. Let, as above, ${\mathbf u}_\nu$ denote the point of intersection between $\Pi_\nu$ and $\Pi_{\nu+1}$
and $\varphi_\nu$ the angle between  $\Pi_\nu$ and $\Pi_{\nu+1}$ at ${\mathbf u}_\nu$. It follows that, up to a sign, $\varphi_\nu$
is identical with the angle between the normals ${\mathbf h}_\nu$ and ${\mathbf h}_{\nu+1}$,
\begin{equation}\label{GP12}
 \left|\varphi_\nu\right|=d({\mathbf h}_\nu,{\mathbf h}_{\nu+1})
 \;.
\end{equation}
Hence the total geodesic curvature of $\Pi$ will be
\begin{equation}\label{GP13}
  \varphi(\Pi)=\sum_{\nu=1}^{N}\varphi_\nu =\pm\sum_{\nu=1}^{N}d({\mathbf h}_\nu,{\mathbf h}_{\nu+1})
  \;,
\end{equation}
the $\pm$ sign depending on the orientation of ${\mathcal S}$.
In the limit $N\rightarrow\infty$ we will, similarly as above, obtain
\begin{equation}\label{GP14}
  \varphi(\Pi)=\sum_{\nu=1}^{N}\varphi_\nu \rightarrow \delta({\mathcal S})=-\int_{0}^{\left|{\mathcal S}\right|}k_g\,ds
  \;.
\end{equation}

The curve ${\mathcal S}$ will be called \textit{simple} iff its geodesic curvature does not change its sign. In this case
we may rewrite the r.~h.~s.~of (\ref{GP14}) in the form
\begin{equation}\label{GP14a}
 -\int_{0}^{\left|{\mathcal S}\right|}k_g\,ds=\pm\left|{\mathcal H}\right|
  \;,
\end{equation}
where we have used that $\sum_{\nu=1}^{N}d({\mathbf h}_\nu,{\mathbf h}_{\nu+1})$ approaches the length of ${\mathcal H}$ for $N\rightarrow\infty$.

The additional results thus obtained are, first, that the parallel transport along ${\mathcal S}$ can be obtained
as a solution of the classical Rabi problem (\ref{G1}) with time $\tau$ replaced by arc length $s$ using the dual magnetic field.
Second, we can geometrically interpret the phase shift $\delta=\delta({\mathcal S})$, not only as the solid angle
${\mathcal A}({\mathcal S})$ (modulo $2\pi$), but also, for simple spin curves,  as the length of the dual loop ${\mathcal H}$, up to a sign, and hence
\begin{equation}\label{GP15}
 \epsilon_g=\frac{\delta}{T}=\pm\frac{\left|{\mathcal H}\right|}{\left|{\mathcal S}\right|}\mbox{ mod }\omega
 \;.
\end{equation}

\subsection{Duality of loops}\label{sec:D}
We start with a loop ${\mathcal S}$ and its dual loop ${\mathcal H}$ on the unit Bloch sphere parametrized
by the arc length $s$ of ${\mathcal S}$ via ${\mathbf s}(s)$ and ${\mathbf h}(s)$ such that
\begin{equation}\label{D1}
 \frac{d}{ds} {\mathbf s}={\mathbf h}\times {\mathbf s}
\end{equation}
and
\begin{equation}\label{D2}
  {\mathbf h}={\mathbf s}\times \frac{d}{ds}{\mathbf s}
\end{equation}
hold. (We will avoid the use of primes for derivatives in this subsection in order to avoid misunderstandings.)
Hence the triple $\left( {\mathbf h},{\mathbf s}, \frac{d}{ds} {\mathbf s}\right)$ will be a right-handed
orthonormal frame for all values of the parameter $s$.
It follows that
\begin{equation}\label{D3}
 \frac{d}{ds}{\mathbf h}={\mathbf s}\times  \frac{d^2}{ds^2}{\mathbf s}
\end{equation}
is orthogonal to ${\mathbf h}$ and ${\mathbf s}$ and hence
\begin{equation}\label{D4}
  \frac{d}{ds} {\mathbf h}=a(s)\, \frac{d}{ds}{\mathbf s}
  \;.
\end{equation}
Moreover,
\begin{equation}\label{D5}
 \frac{d^2}{ds^2} {\mathbf s}= \frac{d}{ds}\left( {\mathbf h}\times{\mathbf s}\right)=
 \left(\frac{d}{ds}{\mathbf h}\right)\times{\mathbf s}+{\mathbf h}\times \left(\frac{d}{ds}{\mathbf s}\right)
 =a\,\left( \frac{d}{ds}{\mathbf s}\right)\times{\mathbf s}-{\mathbf s}=-a\,{\mathbf h}-{\mathbf s}
 \;,
\end{equation}
and hence
\begin{equation}\label{D6}
 g=\left( {\mathbf s}\times \frac{d}{ds}{\mathbf s}\right)\cdot  \frac{d^2}{ds^2}{\mathbf s}=-a\,{\mathbf h}\cdot{\mathbf h}=-a
 \;.
\end{equation}

Now (\ref{D1}), (\ref{D4}) and (\ref{D6}) imply
\begin{equation}\label{D7}
  \frac{d}{ds}{\mathbf h}=-g \frac{d}{ds}{\mathbf s}=\left( g\,{\mathbf s}\right)\times{\mathbf h}\equiv \tilde{\mathbf s}\times{\mathbf h}
  \;.
\end{equation}
The latter equation has the form of (\ref{G1}) and hence can be interpreted in such a way that the ``spin vector" ${\mathbf h}$
moves according to (\ref{G1}) under the influence of the ``magnetic field" $\tilde{\mathbf s}$. In this sense the r\^{o}le of
classical spin and magnetic field is interchanged. However, in general $\tilde{\mathbf s}$ will not be a unit vector
and $s$ will not be the arc length of the loop ${\mathcal H}$.

The situation will be more symmetric if we additionally pass from $s$ to the arc length parameter of ${\mathcal H}$, denoted by $r$.
Then the equation of motion for ${\mathbf h}(r)$ assumes the form
\begin{equation}\label{D8}
  \frac{d}{dr}{\mathbf h} = \frac{ds}{dr} \frac{d}{ds}{\mathbf h}\stackrel{(\ref{D7})}{=}
  \left(\frac{ds}{dr} \, g\,{\mathbf s}\right)\times{\mathbf h}\equiv \overline{\mathbf s}\times{\mathbf h}
  \;.
\end{equation}
Now the new ``magnetic field" $\overline{\mathbf s}$ has unit length since it is the vector product of two
orthogonal unit vectors:
\begin{equation}\label{D9}
 {\mathbf h}\times\frac{d}{dr}{\mathbf h} = {\mathbf h}\times\left(\overline{\mathbf s}\times{\mathbf h}\right)=
 \overline{\mathbf s}\,\underbrace{{\mathbf h}\cdot {\mathbf h}}_1-{\mathbf h}\,\underbrace{\overline{\mathbf s}\cdot {\mathbf h}}_0=\overline{\mathbf s}
 \;.
\end{equation}
This means
\begin{equation}\label{D10}
  \left\|\overline{\mathbf s}\right\|=\left\|\frac{ds}{dr} \, g\,{\mathbf s}\right\|=\left\|{\mathbf s}\right\|=1
  \;,
\end{equation}
and hence
\begin{equation}\label{D11}
 \frac{ds}{dr}=\left| \frac{1}{g}\right|
 \;.
\end{equation}
Together with (\ref{D9}) this implies
\begin{equation}\label{D11a}
  \overline{\mathbf s}(r)=\pm {\mathbf s}(r)
  \;.
\end{equation}

If the r\^{o}le of ${\mathcal S}$ and  ${\mathcal H}$ is interchanged, we obtain
\begin{equation}\label{D11b}
 \frac{dr}{ds}=\left| g\right|=\left|  \frac{1}{G}\right|
 \;,
\end{equation}
where $G$ is the geodesic curvature of ${\mathcal H}$ :
\begin{equation}\label{D11c}
G\equiv {\mathbf h}\cdot\left( \frac{d{\mathbf h}}{dr}\times \frac{d^2{\mathbf h}}{dr^2}\right)
\;.
\end{equation}

Summarizing, we have two loops ${\mathcal S}$ and ${\mathcal H}$ on the unit Bloch sphere that give rise to
two different solutions of (\ref{G1}): Either ${\mathcal S}$ consists of spin vectors and ${\mathcal H}$ of magnetic
field vectors and the time parameter $\tau$ in (\ref{G1}) is chosen as the arc length $s$ of ${\mathcal S}$.
Or, ${\mathcal H}$ consists of spin vectors and $\pm {\mathcal S}$ of magnetic
field vectors and the time parameter $\tau$ in (\ref{G1}) is chosen as the arc length $r$ of ${\mathcal H}$.
This symmetry between ${\mathcal S}$ and ${\mathcal H}$ justifies the denotation as ``dual loops".
The possible sign change indicated by $\pm {\mathcal S}$ signifies that the proper mathematical framework
would be to consider loops, not on the Bloch sphere $S^2$, but in the \textit{projective plane} ${\mathbf P}_2({\mathbbm R})$
obtained by identifying anti-podal points of $S^2$.

For both realizations of solutions of (\ref{G1}) we can calculate the quasienergy denoted by
$\epsilon_S$ and $\epsilon_H$, resp.~. It consist only of its geometric part since spin vector and magnetic field
will be orthogonal in both realizations. For the case of simple spin curves, (\ref{GP15}) immediately implies
that there are representatives of the classes $[\epsilon_S ]$ and  $[\epsilon_H ]$, resp.~, such that
\begin{equation}\label{D12}
  \epsilon_S =\pm\frac{1}{\epsilon_H}
  \;,
\end{equation}
further illustrating the duality between ${\mathcal S}$ and ${\mathcal H}$.\\

\begin{figure}[t]
\centering
\includegraphics[width=0.8\linewidth]{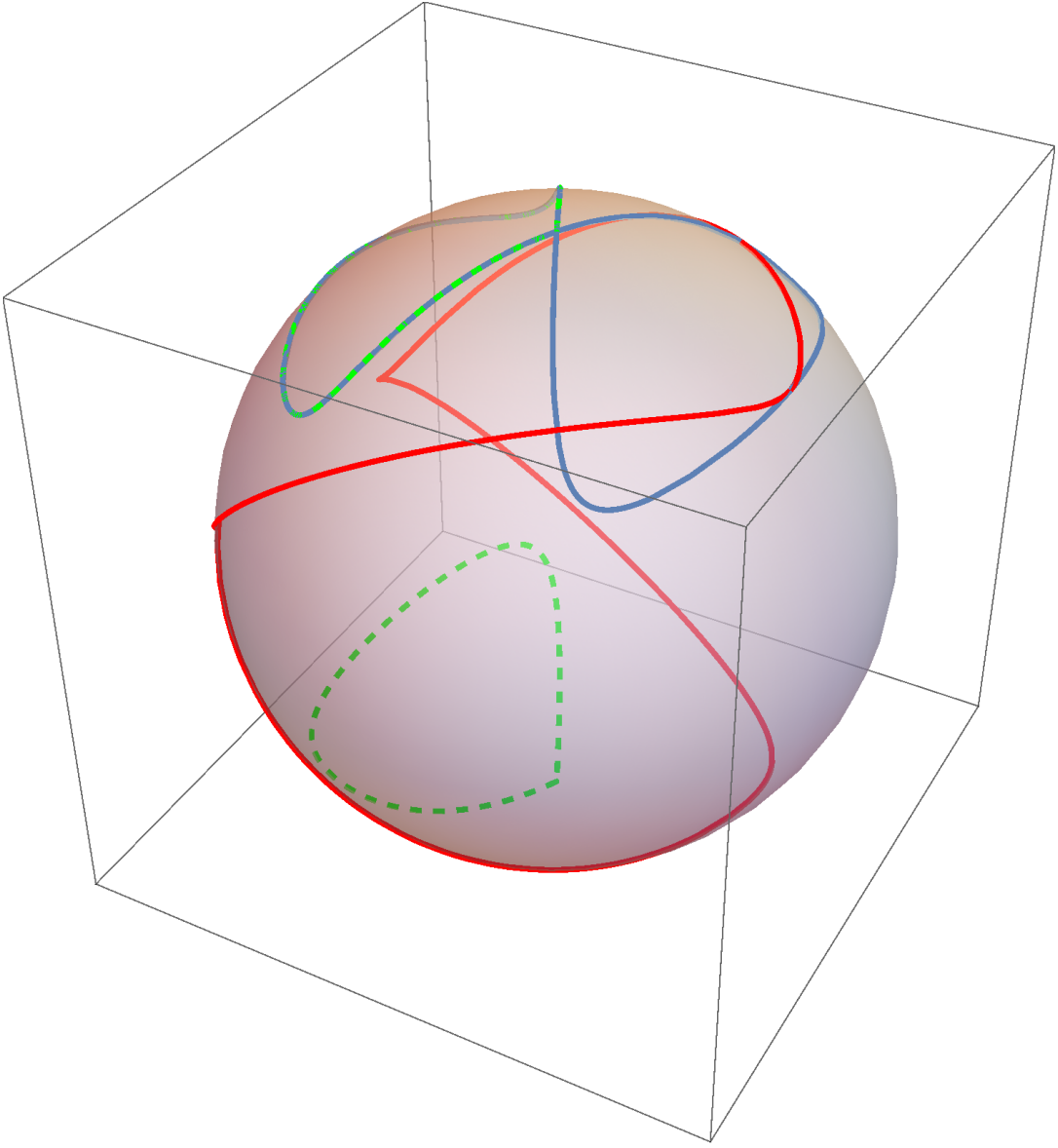}
\caption{Illustration of example $2$ for duality of loops. The blue curve represents the orbit ${\mathcal S}$
of a time-dependent spin function ${\mathbf s}(\tau)$ according to (\ref{D23}); the red one is the dual loop of ${\mathcal H}$ of magnetic field vectors.
If we iterate the construction we obtain for the bi-dual the two green dashed curves ${\mathcal R}$ that locally coincide with  ${\mathcal S}$  or  $-{\mathcal S}$.
}
\label{FIGDD}
\end{figure}

\begin{figure}[t]
\centering
\includegraphics[width=0.8\linewidth]{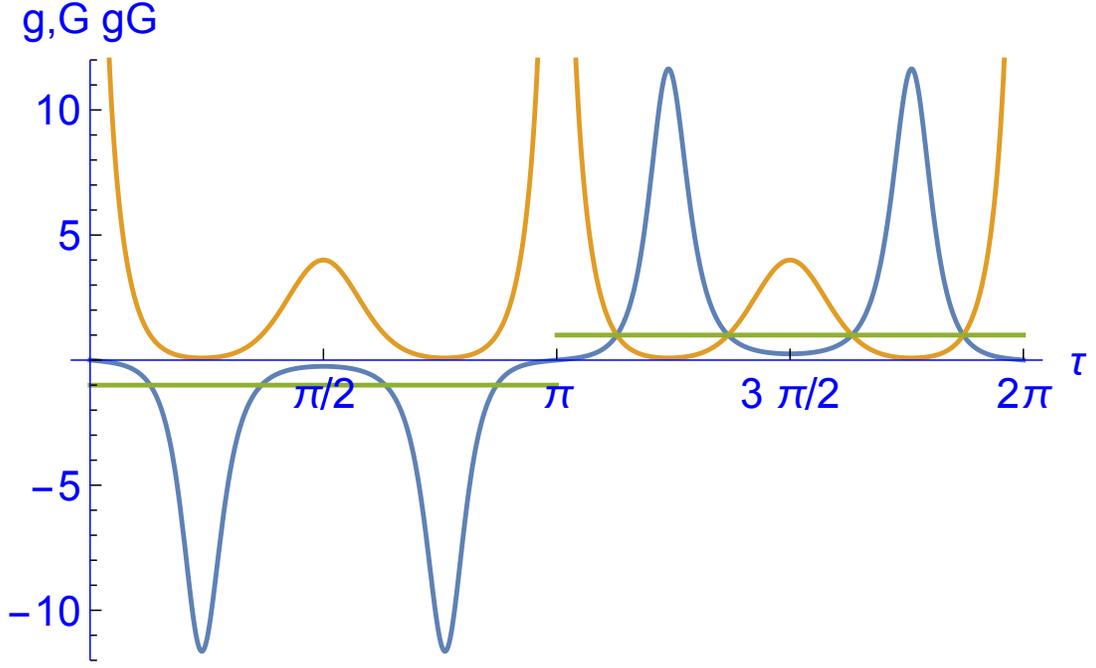}
\caption{Plot of the geodesic curvature $g$ of  ${\mathcal S}$  (blue curve), $G$ of  ${\mathcal H}$  (orange curve),
and the product $g G$ (green curve) as a function of $\tau$. The data are the same as for Figure \ref{FIGDD}.
According to (\ref{D11b}) the product  $g G$ must have the absolute value $1$.
}
\label{FIGG1}
\end{figure}

\subsubsection{Example 2}\label{sec:ex2}

In order to illustrate the notion of duality considered in this subsection we consider two examples.
The first one is a special case of the Rabi problem with circular polarization.
Let
\begin{equation}\label{D13}
  {\mathbf s}(s)=\left(
\begin{array}{c}
 \sqrt{1-z^2} \cos \left(\frac{s}{\sqrt{1-z^2}}\right) \\
 \sqrt{1-z^2} \sin \left(\frac{s}{\sqrt{1-z^2}}\right) \\
 z \\
\end{array}
\right)
\;,
\end{equation}
be the arc length parametrization of a circle ${\mathcal S}$ on $S^2$ lying in the plane $z=\mbox{const.}$ with $-1<z<1$
and $z\neq 0$.
This leads to
\begin{equation}\label{D13a}
\left|{\mathcal S}\right| =2\pi \sqrt{1-z^2}\equiv \frac{2\pi}{\omega}
\;.
\end{equation}
Then (\ref{D2}) yields the parametrization of the dual loop ${\mathcal H}$:
\begin{equation}\label{D14}
  {\mathbf h}(s)=\left(
\begin{array}{c}
 -z \cos \left(\frac{s}{\sqrt{1-z^2}}\right) \\
 -z \sin \left(\frac{s}{\sqrt{1-z^2}}\right) \\
 \sqrt{1-z^2} \\
\end{array}
\right)
\;,
\end{equation}
satisfying
\begin{equation}\label{D14a}
 \left| {\mathcal H}\right|=2\,\pi\,|z|\equiv \frac{2\pi}{\Omega}
 \;.
\end{equation}
After some elementary calculations we obtain
\begin{equation}\label{D15}
g=\left( {\mathbf s}\times \frac{d}{ds}{\mathbf s}\right)\cdot  \frac{d^2}{ds^2}{\mathbf s}=\frac{z}{\sqrt{1-z^2}}
\;,
\end{equation}
and hence the first expression for the quasienergy $\epsilon_S$ reads
\begin{equation}\label{D16}
  \epsilon_S^{(1)}\stackrel{(\ref{QI22})}{=}\frac{1}{\left|{\mathcal S}\right|}\int_{0}^{\left|{\mathcal S}\right|}(-g)\,ds=-g=-\frac{z}{\sqrt{1-z^2}}
  \;.
\end{equation}
The solid angle enclosed by ${\mathcal S}$ will be
\begin{equation}\label{D17}
 {\mathcal A}\left( {\mathcal S}\right)=2\pi\,(1-z)\stackrel{(\ref{D14a})}{=}2\pi\pm\left| {\mathcal H}\right|
 \;,
\end{equation}
and hence a second expression for $\epsilon_S$  will be
\begin{equation}\label{D18}
   \epsilon_S^{(2)}\stackrel{(\ref{QI23})}{=}\frac{ {\mathcal A}\left( {\mathcal S}\right)}{\left|{\mathcal S}\right|}=\frac{\sqrt{1-z}}{\sqrt{1+z}}
   \;.
\end{equation}
It satisfies
\begin{equation}\label{D19}
  \epsilon_S^{(2)}= \epsilon_S^{(1)}+\omega
  \;,
\end{equation}
and hence both expressions (\ref{D16}) and (\ref{D18}) agree modulo $\omega$.

The arc length $r$ corresponding to ${\mathcal H}$ is obtained as
\begin{equation}\label{D20}
 r\stackrel{(\ref{D11})}{=}g\, s \stackrel{(\ref{D15})}{=} \frac{z}{\sqrt{1-z^2}}\,s
 \;.
\end{equation}
After some elementary calculations we obtain
\begin{equation}\label{D21}
G\equiv\left( {\mathbf h}\times \frac{d}{dr}{\mathbf h}\right)\cdot  \frac{d^2}{dr^2}{\mathbf h}=\frac{\sqrt{1-z^2}}{z}=\frac{1}{g}
\;,
\end{equation}
and hence the first expression for the quasienergy $\epsilon_H$ reads
\begin{equation}\label{D22}
  \epsilon_H^{(1)}\stackrel{(\ref{QI22})}{=}\frac{1}{\left|{\mathcal H}\right|}\int_{0}^{\left|{\mathcal H}\right|}(-G)\,dr=-G=-\frac{\sqrt{1-z^2}}{z}
  \;.
\end{equation}
The solid angle enclosed by ${\mathcal H}$ will be
\begin{equation}\label{D23}
 {\mathcal A}\left( {\mathcal H}\right)=2\pi\,(1-\sqrt{1-z^2})\stackrel{(\ref{D13a})}{=}2\pi-\left| {\mathcal S}\right|
 \;,
\end{equation}
and hence a second expression for $\epsilon_S$  will be
\begin{equation}\label{D24}
   \epsilon_H^{(2)}\stackrel{(\ref{QI23})}{=}\frac{ {\mathcal A}\left( {\mathcal H}\right)}{\left|{\mathcal H}\right|}=\frac{1-\sqrt{1-z^2}}{\left| z\right| }
   \;.
\end{equation}
It satisfies
\begin{equation}\label{D25}
  \epsilon_H^{(2)}= \pm\epsilon_H^{(1)}+\Omega
  \;,
\end{equation}
the $\pm$ sign depending on the sign of $z$.
Hence both expressions (\ref{D16}) and (\ref{D18}) agree up to a sign and modulo $\Omega$.

(\ref{D12}) holds for the present example since the triple products $g$ and $G$ are constant and, due to (\ref{D21}), inverses of each other.

In the case of $0<z<1$ the curve ${\mathcal S}$ of our example generates a closed, convex cone ${\mathcal C}\left( {\mathcal S}\right)\subset {\mathbbm R}^3$,
analogously for the dual curve ${\mathcal H}$. Then it follows that ${\mathcal C}\left( {\mathcal H}\right)$ is the \textit{dual cone} of
${\mathcal C}\left( {\mathcal S}\right)$ and vice versa. Here the dual cone $C'$  of a cone $C$ is defined by
\begin{equation}\label{D19a}
  C' \equiv \{{\mathbf x}\in{\mathbbm R}^3\,\left|\,{\mathbf x}\cdot {\mathbf y}\ge 0 \mbox{ for all }{\mathbf y}\in C\right.\}
  \;,
\end{equation}
see, e.~g.~,  \cite{R97}.
In this sense our definition of dual curves is compatible with the established notion of dual cones in ${\mathbbm R}^3$.

Finally, we note that the magnetic field (\ref{D14}) can be understood as a special case of the Rabi problem with circularly polarized
driving (\ref{P11})
if we set $F=-z$, $\omega=\frac{1}{\sqrt{1-z^2}}$ and $\omega_0={\sqrt{1-z^2}}$.
It is well-known \cite{R37}, \cite{S18} that for this problem the quasienergy $\epsilon_c$ assumes the form
\begin{equation}\label{D27}
 \epsilon_c=\Omega_{Rabi}\equiv \sqrt{F^2+(\omega_0-\omega)^2}
 \;,
\end{equation}
taking into account that the quasienergy of the classical Rabi problem is twice the quasienergy of the $s=1/2$ quantum Rabi problem
modulo $\omega$, see also (\ref{P17}).
In our case it follows that
\begin{equation}\label{D28}
 \Omega_{Rabi}= \sqrt{F^2+(\omega-\omega_0)^2}= \sqrt{z^2+\left(\frac{1}{\sqrt{1-z^2}}-{\sqrt{1-z^2}}\right)^2}
 =\frac{|z|}{\sqrt{1-z^2}}
 \;,
\end{equation}
which agrees with (\ref{D16}) up to a possible sign.\\

\subsubsection{Example 3}\label{sec:ex3}

For the second example we take a case where ${\mathcal S}$ is not simple, but of the form of the figure ``$8$" with a double point.
This example also illustrates that we need not explicitly calculate the arc length parameters $s$ of ${\mathcal S}$ or $r$ of ${\mathcal H}$
but may work within the initial parametrization. Let
\begin{equation}\label{ex21}
  {\mathbf s}(\tau)=\frac{1}{\sqrt{\sin ^2(\tau)+\sin ^2(2 \tau)+1}}
  \left(
\begin{array}{c}
 \sin \,\tau  \\
 \sin \,2\, \tau \\
 1 \\
\end{array}
\right).
\end{equation}
We calculate the dual loop ${\mathcal H}\in S^2$ by
\begin{equation}\label{ex22}
  h(\tau) =\frac{1}{\left\|\frac{d{\mathbf s}}{d\tau} \right\|}\,{\mathbf s}\times \frac{d{\mathbf s}}{d\tau}
  \;,
\end{equation}
analogously to (\ref{D2}) but without directly using the arc length parameter $s$. This and the following expressions
can be easily obtained by a computer algebra software but are too involved to be displayed here.
The loop ${\mathcal H}$ is displayed in Figure \ref{FIGDD}. It shows two cusps corresponding to the double point of ${\mathcal S}$.
These cusps necessarily occur according to the following reasoning: At the double point
corresponding to the values $\tau=0,\pi$ of the parameter,
the geodesic curvature $g$ of ${\mathcal S}$ changes its sign, see Figure \ref{FIGG1}. According to (\ref{D11a})
the geodesic curvature $G$ of ${\mathcal H}$ must diverge at $\tau=0,\pi$ which explains the two cusps.

The curve ${\mathcal S}$ can be divided into the parts ${\mathcal S}_1$ and ${\mathcal S}_2$ that have in common
only the double point. The corresponding parts of ${\mathcal H}$ are denoted by ${\mathcal H}_i,\;i=1,2$.
Both parts ${\mathcal S}_i$ encircle solid angles ${\mathcal A}\left( {\mathcal S}_i\right)\approx \pm 4.64172$
that correspond to the length of the ${\mathcal H}_i$.
But due to the different signs the total solid angle and the corresponding quasienergy vanishes.

Interestingly, if we calculate the ``bi-dual" loop ${\mathcal R}$ according to
\begin{equation}\label{ex23}
  r(\tau) =\frac{1}{\left\|\frac{d{\mathbf h}}{d\tau} \right\|}\,{\mathbf h}\times \frac{d{\mathbf h}}{d\tau}
  \;,
\end{equation}
then we obtain two disjoint curves that locally coincide with  $-{\mathcal S}_1$ and  ${\mathcal S}_2$, see Figure \ref{FIGDD}.

\subsection{New solutions}\label{sec:N}

In this subsection we will show how the results of the preceding subsection \ref{sec:D}
can be used to generate new solutions of (\ref{G1}).
We start with two dual loops ${\mathcal S}$ and ${\mathcal H}$ on the unit Bloch sphere parametrized
by the arc length $s$ of ${\mathcal S}$  such that the equations (\ref{D1}) -- (\ref{D11}) of subsection \ref{sec:D} are satisfied.

Then we consider a second curve $\tilde{\mathcal S}\subset S^2$ parametrized by
\begin{equation}\label{N7}
 \tilde{\mathbf s}(s)=\cos\alpha(s)\, {\mathbf s}(s) +\sin\alpha(s)\,{\mathbf h}(s)
 \;,
\end{equation}
where $s\mapsto \alpha(s)$ is an arbitrary smooth function. Solving (\ref{N7}) for ${\mathbf s}$ gives
\begin{equation}\label{N8}
 {\mathbf s}= (\sec\alpha)\, \tilde{\mathbf s}-(\tan\alpha)\,{\mathbf h}
 \;.
\end{equation}
We want to derive an equation of motion for $\tilde{\mathbf s}$ and consider the following equations:
\begin{eqnarray}
\label{N9a}
  \frac{d}{ds}\,\tilde{\mathbf s} &=& (\cos\alpha)\,{\mathbf s}'+(\sin\alpha)\,{\mathbf h}'
  +\alpha'\,\left(-(\sin\alpha)\,{\mathbf s}+(\cos\alpha)\,{\mathbf h} \right) \\
  \label{N9b}
  &\stackrel{(\ref{D1},\ref{D4},\ref{D6})}{=}&(\cos\alpha)\,{\mathbf h}\times{\mathbf s}-g\,(\sin\alpha)\,{\mathbf h}\times{\mathbf s}+
  \alpha'\,\left(-(\sin\alpha)\,{\mathbf s}+(\cos\alpha)\,{\mathbf h} \right) \\
  \label{N9c}
  &\stackrel{(\ref{N8})}{=}&(1-g\,\tan\alpha)\,{\mathbf h}\times \tilde{\mathbf s}+
  \alpha'\;\left(-(\sin\alpha)\,{\mathbf s}+(\cos\alpha)\,{\mathbf h}\right)
  \equiv (1-g\,\tan\alpha)\,{\mathbf h}\times \tilde{\mathbf s}+{\mathbf v}
  \;.
\end{eqnarray}
The vector ${\mathbf v}$ will be orthogonal to $\tilde{\mathbf s}$ by
\begin{equation}\label{N10}
  {\mathbf v}\cdot \tilde{\mathbf s}=
  \left( \alpha'\;\left(-(\sin\alpha)\,{\mathbf s}+(\cos\alpha)\,{\mathbf h}\right)\right)
  \cdot\left((\cos\alpha)\,{\mathbf s}+(\sin\alpha)\,{\mathbf h} \right)
  =(\cos\alpha)(\sin\alpha)(-\alpha'+\alpha')=0
  \;.
\end{equation}
Using $\tilde{\mathbf s}\cdot\tilde{\mathbf s}=1$ we obtain
\begin{eqnarray}\label{N11}
 \left( \tilde{\mathbf s}\times {\mathbf v}\right)\times \tilde{\mathbf s}&\stackrel{(\ref{N10})}{=}&
 \underbrace{\left( \tilde{\mathbf s}\cdot\tilde{\mathbf s}\right)}_1 {\mathbf v}
 -\underbrace{\left({\mathbf v}\cdot\tilde{\mathbf s}\right)}_0\tilde{\mathbf s}= {\mathbf v}
 \;,
\end{eqnarray}
and
\begin{equation}\label{N12}
   \tilde{\mathbf s}\times {\mathbf v}=\left((\cos\alpha)\,{\mathbf s}+(\sin\alpha)\,{\mathbf h} \right)\times
   \left( \alpha'\;\left(-(\sin\alpha)\,{\mathbf s}+(\cos\alpha)\,{\mathbf h}\right)\right)=
   -\alpha'\left((\cos\alpha)^2+(\sin\alpha)^2\right)\,{\mathbf h}\times{\mathbf s}= -\alpha'\,{\mathbf h}\times{\mathbf s}
   \;.
\end{equation}
Finally,
\begin{equation}\label{N13}
\frac{d}{ds}\tilde{\mathbf s}=\left((1-g\,\tan\alpha){\mathbf h}-\alpha'\,{\mathbf h}\times{\mathbf s} \right)\times \tilde{\mathbf s}
\equiv \tilde{\mathbf h}\times\tilde{\mathbf s}
\;,
\end{equation}
and hence $\tilde{\mathbf s}$ satisfies a Rabi type equation of motion with a modified magnetic field $\tilde{\mathbf h}$.
Upon a suitable choice of $\alpha(s)$ the new solution $\tilde{\mathbf s}$ will also be periodic in the original arc length parameter $s$ of
${\mathcal S}$.

\subsubsection{Example 4}\label{ex4}

\begin{figure}[t]
\centering
\includegraphics[width=0.7\linewidth]{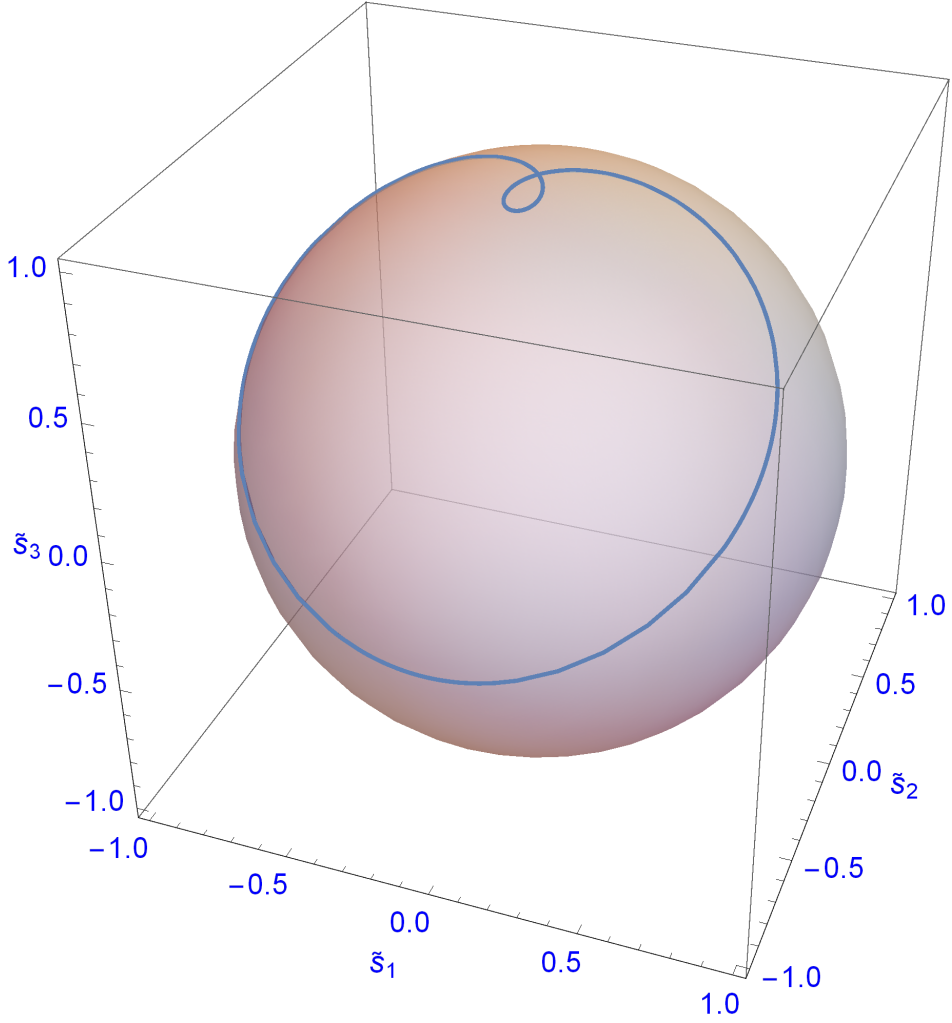}
\caption{Illustration of the loop of spin vectors $\tilde{\mathbf s}$ on the Bloch sphere corresponding to (\ref{N17}).
The parameter $z$ has been chosen as $z=\frac{1}{\sqrt{2}}$.
}
\label{FIGNS1}
\end{figure}

\begin{figure}[t]
\centering
\includegraphics[width=0.7\linewidth]{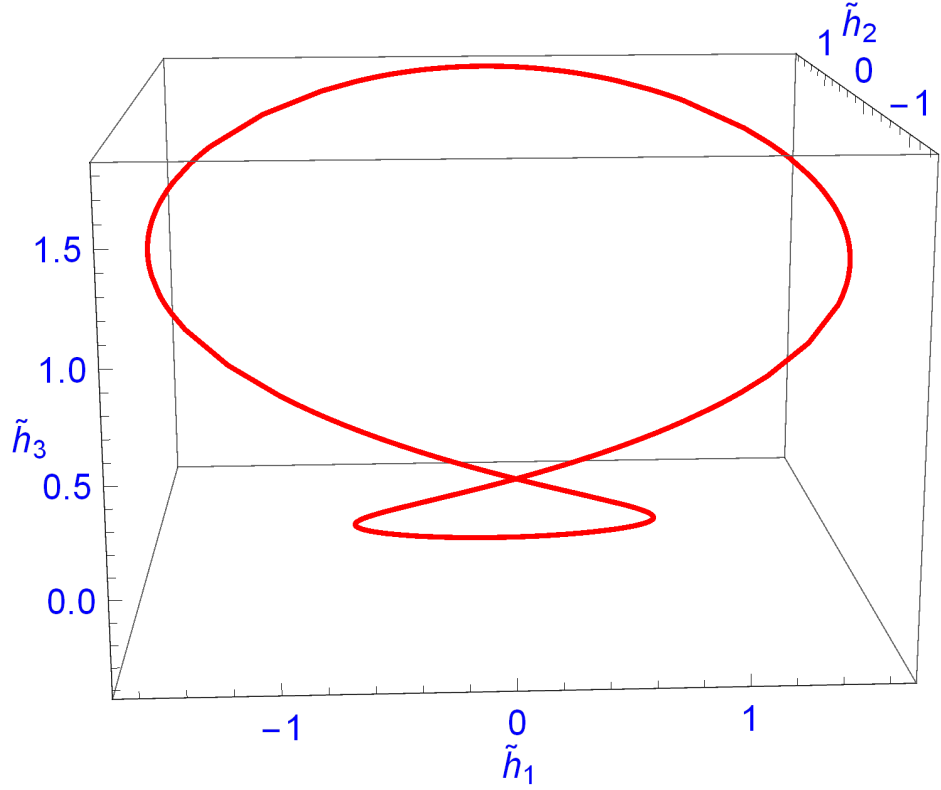}
\caption{Illustration of the loop of magnetic field vectors $\tilde{\mathbf h}$  corresponding to (\ref{N18}).
The parameter $z$ has been chosen as $z=\frac{1}{\sqrt{2}}$.
}
\label{FIGNH1}
\end{figure}

As an example we choose ${\mathbf s}(s)$ according to (\ref{D13}) and $ {\mathbf h}(s)$ according to (\ref{D14}), further
\begin{equation}\label{N16}
  \alpha(s)=\sin \left(\frac{s}{\sqrt{1-z^2}}\right)
  \;.
\end{equation}
This entails the new solution of (\ref{N13}) of the form
\begin{equation}\label{N17}
 \tilde{\mathbf s}(s)=\left(
\begin{array}{c}
 \cos \left(\frac{s}{\sqrt{1-z^2}}\right) \left(\sqrt{1-z^2} \cos \left(\sin
   \left(\frac{s}{\sqrt{1-z^2}}\right)\right)-z \sin \left(\sin
   \left(\frac{s}{\sqrt{1-z^2}}\right)\right)\right) \\
 \sin \left(\frac{s}{\sqrt{1-z^2}}\right) \left(\sqrt{1-z^2} \cos \left(\sin
   \left(\frac{s}{\sqrt{1-z^2}}\right)\right)-z \sin \left(\sin
   \left(\frac{s}{\sqrt{1-z^2}}\right)\right)\right) \\
 \sqrt{1-z^2} \sin \left(\sin \left(\frac{s}{\sqrt{1-z^2}}\right)\right)+z \cos
   \left(\sin \left(\frac{s}{\sqrt{1-z^2}}\right)\right) \\
\end{array}
\right)
\;,
\end{equation}
with
\begin{equation}\label{N18}
 \tilde{\mathbf h}(s)=\left(
\begin{array}{c}
 \frac{\cos \left(\frac{s}{\sqrt{1-z^2}}\right) \left(\sin
   \left(\frac{s}{\sqrt{1-z^2}}\right)+z \left(z \tan \left(\sin
   \left(\frac{s}{\sqrt{1-z^2}}\right)\right)-\sqrt{1-z^2}\right)\right)}{\sqrt{1-
   z^2}} \\
 \frac{z \sin \left(\frac{s}{\sqrt{1-z^2}}\right) \left(z \tan \left(\sin
   \left(\frac{s}{\sqrt{1-z^2}}\right)\right)-\sqrt{1-z^2}\right)-\cos
   ^2\left(\frac{s}{\sqrt{1-z^2}}\right)}{\sqrt{1-z^2}} \\
 \sqrt{1-z^2}-z \tan \left(\sin \left(\frac{s}{\sqrt{1-z^2}}\right)\right) \\
\end{array}
\right)
\;,
\end{equation}
see Figures \ref{FIGNS1} and \ref{FIGNH1}, where $z$ has been chosen as $z=\frac{1}{\sqrt{2}}$.

\subsection{Quasienergy II}\label{sec:QII}

In this subsection we are going to show that our result (\ref{QI15}) for the quasienergy of a periodic solution
of (\ref{G1}) is equivalent to the integral obtained in \cite{S18}. The corresponding calculations are elementary but
somewhat lengthy.
Let  ${\mathbf s}(\tau)$  and ${\mathbf r}(\tau)$ be two solutions of (\ref{G1})
such that ${\mathbf h}(\tau)$ and ${\mathbf s}(\tau)$ are  $T$-periodic and ${\mathbf r}(0)\cdot{\mathbf s}(0)=0$ and hence
\begin{equation}\label{QII1}
 {\mathbf r}(\tau)\cdot{\mathbf s}(\tau)=0 \mbox{ for all } \tau\in{\mathbbm R}
 \;.
\end{equation}
Further, let ${\mathbf f}$ be a constant unit vector that is not contained in the curve ${\mathcal S}$.
We consider the right-handed orthogonal, not necessarily normalized,  ``${\mathbf f}$-frame"
\begin{equation}\label{QII2}
\left( {\mathbf s}(\tau),{\mathbf f}\times{\mathbf s}(\tau),{\mathbf s}(\tau)\times\left({\mathbf f}\times{\mathbf s}(\tau)\right)\right)
\;,
\end{equation}
and expand ${\mathbf r}(\tau)$ w.~r.~t.~this frame:
\begin{equation}\label{QII3}
{\mathbf r}(\tau)=a(\tau)\,{\mathbf f}\times{\mathbf s}(\tau)+b(\tau)\,{\mathbf s}(\tau)\times\left({\mathbf f}\times{\mathbf s}(\tau)\right)
\;,
\end{equation}
taking into account (\ref{QII1}). Differentiating (\ref{QII3}) w.~r.~t.~time,
using the equation of motion (\ref{G1})  and expanding the multiple vector products gives
\begin{eqnarray}
\label{QII4a}
  \dot{\mathbf r} &=& \dot{a}\,{\mathbf f}\times{\mathbf s}+\dot{b}\,{\mathbf s}\times\left({\mathbf f}\times{\mathbf s}\right)
  + a\, {\mathbf f}\times\dot{\mathbf s}+b\left(\dot{\mathbf s}\times\left({\mathbf f}\times{\mathbf s}\right)+
  {\mathbf s}\times\left({\mathbf f}\times\dot{\mathbf s}\right) \right)\\
  \label{QII4b}
  &=& \dot{a}\,{\mathbf f}\times{\mathbf s}+\dot{b}\,{\mathbf s}\times\left({\mathbf f}\times{\mathbf s}\right)
  + a\, {\mathbf f}\times({\mathbf h}\times{\mathbf s})+b\left(({\mathbf h}\times{\mathbf s})\times\left({\mathbf f}\times{\mathbf s}\right)+
  {\mathbf s}\times\left({\mathbf f}\times({\mathbf h}\times{\mathbf s})\right) \right)\\
  \label{QII4c}
  &=&\dot{a}\,{\mathbf f}\times{\mathbf s}+\dot{b}\,{\mathbf s}\times\left({\mathbf f}\times{\mathbf s}\right)
  +a\left( {\mathbf f}\cdot{\mathbf s}\;{\mathbf h}- {\mathbf f}\cdot{\mathbf h}\;{\mathbf s}\right)
  +b\left({\mathbf h}\cdot({\mathbf f}\times{\mathbf s})\;{\mathbf s}+ {\mathbf f}\cdot{\mathbf s}\;{\mathbf s}\times{\mathbf h}\right)
  \;.
\end{eqnarray}
Another expression for $ \dot{\mathbf r}$ is obtained by inserting (\ref{QII3}) into the equation of motion:
\begin{eqnarray}
\label{QII5a}
   \dot{\mathbf r} &=&  {\mathbf h}\times{\mathbf r}
   ={\mathbf h}\times\left(a\;{\mathbf f}\times{\mathbf s}+b\; {\mathbf s}\times\left({\mathbf f}\times{\mathbf s}\right)\right) \\
   \label{QII5b}
   &=& a\left({\mathbf h}\cdot{\mathbf s}\;{\mathbf f}- {\mathbf h}\cdot{\mathbf f}\;{\mathbf s}\right)+
   b\left( {\mathbf h}\times{\mathbf f}-{\mathbf f}\cdot{\mathbf s}\;{\mathbf h}\times{\mathbf s}\right)
   \;.
\end{eqnarray}
We will expand (\ref{QII4c}) and (\ref{QII5b}) w.~r.~t.~the ${\mathbf f}$-frame. The ${\mathbf s}$-components give no new results.
For the ${\mathbf f}\times{\mathbf s}$-components we obtain
\begin{eqnarray}
\label{QII6a}
  \dot{\mathbf r}\cdot({\mathbf f}\times{\mathbf s}) &\stackrel{(\ref{QII4c})}{=}&
  \dot{a}\;({\mathbf f}\times{\mathbf s})\cdot({\mathbf f}\times{\mathbf s})+
  a\;{\mathbf f}\cdot{\mathbf s}\;{\mathbf h}\cdot({\mathbf f}\times{\mathbf s})+
  b\;{\mathbf f}\cdot{\mathbf s}\;({\mathbf s}\times{\mathbf h})\times({\mathbf f}\times{\mathbf s})   \\
   \label{QII6b}
   &=&\dot{a}\;\left(1-({\mathbf f}\cdot{\mathbf s})^2\right)+
   a\;{\mathbf f}\cdot{\mathbf s}\;{\mathbf h}\cdot({\mathbf f}\times{\mathbf s})+
   b\;{\mathbf f}\cdot{\mathbf s}\;({\mathbf s}\times{\mathbf h})\times({\mathbf f}\times{\mathbf s}) \\
   \label{QII6c}
   &\stackrel{(\ref{QII5b})}{=}&
   b\;\left( ({\mathbf h}\times{\mathbf f})\cdot({\mathbf f}\times{\mathbf s})-{\mathbf f}\cdot{\mathbf s}\;
   ({\mathbf h}\times{\mathbf s})\cdot({\mathbf f}\times{\mathbf s})\right)\\
    \label{QII6d}
    &=& b\;\left({\mathbf f}\cdot{\mathbf h}\;{\mathbf f}\cdot{\mathbf s}-{\mathbf s}\cdot{\mathbf h}-{\mathbf f}\cdot{\mathbf s}\;
   ({\mathbf h}\times{\mathbf s})\cdot({\mathbf f}\times{\mathbf s})\right)
   \;.
\end{eqnarray}
Subtracting (\ref{QII6d}) from  (\ref{QII6b}) yields
\begin{equation}\label{QII7}
 \dot{a}\;\left(1-({\mathbf f}\cdot{\mathbf s})^2\right)=
 -a\;{\mathbf f}\cdot{\mathbf s}\;{\mathbf h}\cdot({\mathbf f}\times{\mathbf s})
 -b\;\left({\mathbf s}\cdot{\mathbf h}-{\mathbf f}\cdot{\mathbf h}\;{\mathbf f}\cdot{\mathbf s}\right)
 \;.
\end{equation}
The analogous calculation of the ${\mathbf s }\times ({\mathbf f}\times{\mathbf s})$-components that will not be given in detail yields
\begin{equation}\label{QII8}
  \dot{b}\;\left(1-({\mathbf f}\cdot{\mathbf s})^2\right)=
 -b\;{\mathbf f}\cdot{\mathbf s}\;{\mathbf h}\cdot({\mathbf f}\times{\mathbf s})
 +a\;\left({\mathbf s}\cdot{\mathbf h}-{\mathbf f}\cdot{\mathbf h}\;{\mathbf f}\cdot{\mathbf s}\right)
 \;.
\end{equation}
The two equations (\ref{QII7}) and (\ref{QII8}) together imply
\begin{equation}\label{QII9}
  \frac{\dot{b}\,a-\dot{a}\,b}{a^2+b^2}=
  \frac{{\mathbf s}\cdot{\mathbf h}-{\mathbf f}\cdot{\mathbf h}\;{\mathbf f}\cdot{\mathbf s}}{1-({\mathbf f}\cdot{\mathbf s})^2}
  \;.
\end{equation}
Recall that ${\mathbf f}\times{\mathbf s}$ and ${\mathbf s }\times ({\mathbf f}\times{\mathbf s})$ have the same length
hence the angle $\varphi(\tau)$ between ${\mathbf r}(\tau)$ and ${\mathbf f}\times{\mathbf s}(\tau)$
satisfies
\begin{equation}\label{QII10}
  \tan \varphi =\frac{b}{a}\equiv u
  \;,
\end{equation}
which entails
\begin{equation}\label{QII11}
    \frac{\dot{b}\,a-\dot{a}\,b}{a^2+b^2}=\frac{\dot{u}}{1+u^2}=\frac{d}{d\tau}\arctan u=\dot{\varphi}
    \;.
\end{equation}
The total phase shift $\delta$ between ${\mathbf r}(\tau)$ and ${\mathbf f}\times{\mathbf s}(\tau)$
over one period can thus be written as
\begin{equation}\label{QII12}
 \delta=\int_{0}^{T}\dot{\varphi}\,d\tau \stackrel{(\ref{QII9},\ref{QII11})}{=}
 \int_{0}^{T}  \frac{{\mathbf s}\cdot{\mathbf h}-{\mathbf f}\cdot{\mathbf h}\;{\mathbf f}\cdot{\mathbf s}}{1-({\mathbf f}\cdot{\mathbf s})^2}\,d\tau
 \;.
\end{equation}
Consequently, the quasienergy can be written as
\begin{equation}\label{QII13}
 \epsilon=\frac{\delta}{T}
 =\frac{1}{T}\int_{0}^{T}  \frac{{\mathbf s}\cdot{\mathbf h}-{\mathbf f}\cdot{\mathbf h}\;{\mathbf f}\cdot{\mathbf s}}{1-({\mathbf f}\cdot{\mathbf s})^2}\,d\tau
 \;.
\end{equation}
For the comparison with the analogous expression in \cite{S18} we choose
\begin{equation}\label{QII14}
  {\mathbf f}=\left( \begin{array}{c}
                       0 \\
                       0 \\
                       1
                     \end{array}\right)
                     \mbox{ and write }
                     {\mathbf s}=
                     \left( \begin{array}{c}
                       X \\
                       Y \\
                       Z
                     \end{array}\right)
                     \;,
\end{equation}
which yields
\begin{equation}\label{QII15}
\delta=\int_{0}^{T}\frac{h_1\,X+h_2\,Y}{1-Z^2}\,d\tau
\;.
\end{equation}
Let $\phi(\tau)$ be the azimuthal angle of ${\mathbf s}(\tau)$ w.~r.~t.~the constant standard frame such that
\begin{equation}\label{QIII16}
  \dot{\phi}=\frac{X\,\dot{Y}-Y\,\dot{X}}{X^2+Y^2}
  \;.
\end{equation}
The integral $\int_{0}^{T}\dot{\phi}\,d\tau$ will assume the value $2\pi n$ where $n\in{\mathbbm Z}$ is the
winding number of the periodic solution ${\mathbf s}$ around the $3$-axis. Since $\delta$ is only relevant up to
integer multiples of $2\pi$ we may freely add $\dot{\phi}$ to the integrand of (\ref{QII15}) and obtain:
\begin{eqnarray}
\label{QII17a}
  \frac{h_1\,X+h_2\,Y}{1-Z^2}+\dot{\phi} &=& \frac{1}{1-Z^2}
  \left( h_1\,X+h_2\,Y +X\,\dot{Y}-Y\,\dot{X}
  \right)   \\
  \label{QII17b}
     &=& \frac{1}{1-Z^2}  \left( h_1\,X+h_2\,Y +X\,(h_3\,X-h_1\,Z)-Y\,(h_2\,Z-h_3\,Y)
  \right)   \\
  \label{QII17c}
   &=&\frac{1}{1-Z^2}  \left( h_1\,X+h_2\,Y+h_3(X^2+Y^2)-(h_1\,X+h_2\,Y)Z
  \right)   \\
  \label{QII17d}
  &=&h_3+\frac{1}{1-Z^2}  ( h_1\,X+h_2\,Y)(1-Z)     \\
  \label{QII17e}
  &=&h_3+\frac{h_1\,X+h_2\,Y}{1+Z}
  \;,
\end{eqnarray}
 which agrees with Eq.~(62) of \cite{S18} up to a factor of $\frac{1}{2}$ as discussed above.

\section{Summary}\label{S}
The Rabi oscillations of a periodically driven two-level system can be translated into certain
orbits on the Bloch sphere that are solutions of the classical Rabi problem $\dot{\mathbf s}={\mathbf h}\times{\mathbf s}$.
The latter differential equation has various physical and geometric ramifications. It can be analyzed in terms of Floquet theory
leading to the notion of a ``quasienergy" $\epsilon$. The geometric part $\epsilon_g$ of the quasienergy in turn can be related to the
solid angle encircled by the closed curve ${\mathcal S}$ given by the solution ${\mathbf s}(\tau)$ via the theorem of Gauss-Bonnet, or, alternatively,
understood in terms of the geometric phase similarly as for the Foucault pendulum. These well-known connections
are recapitulated in the present paper and illustrated by a couple of examples.
A probably new result is that $\epsilon_g$ can also be expressed through the length of the ``dual loop" ${\mathcal H}$,
and that the r\^{o}le of ${\mathcal S}$ and ${\mathcal H}$ can, in a certain sense, be interchanged.
For the special case of simple curves the quasienergies of a dual pair of curves are reciprocal.
The closer study of this ``duality of loops" is devoted to future papers.

\begin{acknowledgments}
I thank the members of the  DFG Research Unit FOR 2692 for stimulating and
insightful discussions on the topic of this paper.
\end{acknowledgments}

\end{document}